\documentclass[11pt]{article}

\usepackage{a4wide}
\usepackage{graphicx}
\usepackage{amsmath}
\usepackage{slashed}
\usepackage{listings}
\usepackage{color}

\usepackage{cite}
\usepackage{fancyheadings}
\usepackage[title,titletoc]{appendix}

\definecolor{shaded}{RGB}{245,245,245}
\def\preprintn{MPP-2014-50}

\fancypagestyle{first}
{
  
  \fancyhead[R]{\footnotesize \preprintn}
}

\usepackage[colorlinks=true,linkcolor=blue,citecolor=blue]{hyperref}

\newcommand{\comment}[2]{#2}
\def\email#1{\texttt{#1}}
\def\nn{\nonumber}

\def\O{y}

\def\I{\mathcal{I}}
\def\M{\mathcal{M}}
\def\N{\mathcal{N}}
\def\O{\mathcal{O}}

\def\N{\mathcal{N}}

\def\spa#1{\langle#1\rangle}
\def\spb#1{[#1]}

\def\nn{\nonumber \\}

\title{\textbf{{\sc Ninja}: Automated Integrand Reduction \\ via Laurent Expansion \\ for
One-Loop Amplitudes}}

\date{}

\comment{
} 

\author{Tiziano Peraro\\
\emph{Max-Planck Insitut f\"ur Physik,}\\ \emph{F\"ohringer Ring, 6, D-80805 M\"unchen, Germany}\\
\emph{E-mail:} \email{peraro@mppmu.mpg.de}}

\begin{document}

\maketitle

\thispagestyle{first}

\begin{abstract}
  We present the public \textsc{C++} library \textsc{Ninja}, which
  implements the Integrand Reduction via Laurent Expansion method for
  the computation of one-loop integrals.  The algorithm is suited for
  applications to complex one-loop processes.
\end{abstract}

\clearpage

\tableofcontents

\section{Introduction}
\label{sec:introduction}
The investigation of the factorization properties of scattering
amplitudes at their singularities
\cite{Cachazo:2004kj,Britto:2004ap,Britto:2005fq,Bern:1994zx,Britto:2004nc}
lead to important results in Quantum Field Theory and Theoretical
Particle Physics, such as the development of new methods for
phenomenological computations.  In particular, integrand reduction
methods, developed for one-loop
diagrams~\cite{Ossola:2006us,Ellis:2007br} and recently extended to
higher
loops~\cite{Mastrolia:2011pr,Badger:2012dp,Zhang:2012ce,Mastrolia:2012an,Mastrolia:2013kca},
use the knowledge of the analytic and algebraic structure of loop
integrands in order to rewrite scattering amplitudes as linear
combinations of Master Integrals.

At one loop, integrand-reduction methods allow to express any
integrand in dimensional regularization as a sum of contributions with
at most five propagators in the loop, regardless of the number of
external legs of the amplitude.  The numerators of these contributions
are \emph{polynomial residues} which have a universal parametric form
that does not depend on the process.  This parametric form can be
written as a sum of monomials in the components of the loop momentum,
multiplied by unknown process-dependent coefficients.  After
integration, the amplitude becomes a linear combination of known
integrals.  The coefficients of this linear combination can be
identified with a subset of the ones which parametrize the residues.
Therefore, the problem of the computation of any one-loop amplitude
can be reduced to the one of performing a polynomial fit of the
coefficients of the residues.

The fit of the unknown coefficients can be efficiently performed by
evaluating the numerator of the integrand on \emph{multiple cuts},
i.e.\ on values of the loop momentum such that a subset of the loop
denominators vanish.  The multiple-cut conditions can be viewed as
\emph{projectors} which isolate the residue sitting on the cut
denominators.  A residue can be evaluated by putting on-shell the
corresponding loop propagators and subtracting from the integrand the
non-vanishing contributions coming from higher-point residues.  This
leads to a top-down algorithm which allows to compute any one-loop
amplitude with any number of external legs.

Within semi-numerical computations, the algorithm is usually
implemented by sampling the integrand on several solutions of the
multiple cuts and subtracting at each step of the reduction all the
non-vanishing contributions coming from higher-point residues.  This
yields a system of equations for the coefficients of each residue.
The method is suited for automation and it has been implemented in
several codes, some of which are public (e.g.\
\textsc{CutTools}~\cite{Ossola:2007ax} and
\textsc{Samurai}~\cite{Mastrolia:2010nb}).  Its usage within several
automated frameworks
\cite{Hahn:1998yk,vanHameren:2009dr,Bevilacqua:2011xh,Berger:2008sj,Hirschi:2011pa,Cullen:2011ac,Cascioli:2011va,Badger:2010nx,Badger:2012pg}
has been particularly successful and produced highly non-trivial
phenomenological results.

In this paper we present a new public \textsc{C++} library called
\textsc{Ninja}, which implements an alternative integrand-reduction
algorithm first proposed in Ref.~\cite{Mastrolia:2012bu}.  This is
based on the systematic application of the Laurent series expansion to
an integrand on the multiple cuts.  After performing a suitable
Laurent expansion on a multiple cut, in the asymptotic limit both the
integrand and the subtraction terms exhibit the same polynomial
behavior as the residue.  This allows one to directly identify the
coefficients of the residues (and thus the ones of the Master
Integrals) with the ones of the Laurent expansion of the integrand,
corrected by subtractions terms which can be computed once and for all
as functions of a subset of the higher-point coefficients.  This leads
to a diagonal system of equations for each residue and to a
significant reduction of the number of subtraction terms which affect
the computation of lower-point contributions.

\textsc{Ninja} takes as input the numerator cast into four different
forms.  The first is a procedure which evaluates it as a function of
the loop momentum.  The others instead compute the leading terms of
properly defined Laurent expansions of the numerator.
Since the integrand of a one-loop amplitude is a rational function of
the loop momentum, the Laurent expansions for an integrand can be
obtained via a partial fraction decomposition.  \textsc{Ninja}
implements it semi-numerically via a simplified polynomial division
algorithm between the expansions of the numerator and the ones of the
denominators.
The coefficients of the Laurent expansion are then
corrected by the subtraction terms and multiplied by the Master
Integrals.  These are computed by interfacing \textsc{Ninja} with an
external library which can be specified by the user.  Interfaces for
\textsc{OneLoop} \cite{vanHameren:2010cp,vanHameren:2009dr} and {\sc
  LoopTools}~\cite{Hahn:1998yk} are already provided with the
distribution.

The simplified subtractions and the diagonal systems of equations make
the algorithm implemented in \textsc{Ninja} significantly simpler and
lighter than the traditional one.  The library has been interfaced
with the one-loop package \textsc{GoSam}~\cite{Cullen:2011ac} and has
already been used to compute NLO corrections to Higgs boson production
in association with a top-quark pair and a
jet~\cite{vanDeurzen:2013xla} and several six-, seven- and eight-point
amplitudes involving both massive and massless particles as external
states or propagating in the loop~\cite{vanDeurzen:2013saa}.  These
applications showed that \textsc{Ninja} has better performance and
numerical stability than implementations of traditional integrand
reduction algorithms.  In particular, Ref.~\cite{vanDeurzen:2013saa}
provides a detailed analysis of the performance and accuracy of this
library.  With this paper, we make \textsc{Ninja} publicly available
as a standalone library which can be interfaced to other packages and
frameworks for one-loop computations.

In order to simplify the generation of the numerator expansions and
the corresponding source code needed by \textsc{Ninja} as input, we
distribute with the library a small \textsc{Python} package called
\textsc{NinjaNumGen} which uses
\textsc{Form-4}~\cite{Vermaseren:2000nd,Kuipers:2012rf,Kuipers:2013pba}
in order to compute the expansions and produce an optimized source
code which can be used by the library.

In this paper, besides the description of the implementation of the
algorithm and the usage of the library, we also propose a method for
the extension of the integrand reduction via Laurent expansion for the
computation of integrals whose rank is one unit higher than the number
of loop denominators.  This method is not present elsewhere in the
literature and has been implemented in \textsc{Ninja}, allowing thus
to use the library for computations in non-renormalizable and
effective theories as well.

The paper is organized as follows.  In Section~\ref{sec:laurent} we
give a review of the Laurent Expansion method for the integrand
reduction of one-loop amplitudes.  In Section~\ref{sec:ninja} we
discuss its semi-numerical implementation in the library
\textsc{Ninja}.  The usage of the library is explained, with the help
of simple examples, in Section~\ref{sec:usage}.  In
Section~\ref{sec:examples} we give a description of the examples which
are distributed with the library, giving a better view of its usage
and capabilities.  Appendix \ref{sec:ninjanumgen} gives more details
on the usage of the package \textsc{NinjaNumGen}.  In Appendix
\ref{sec:hr} we present the extension of the algorithm to higher-rank
integrals.  Appendix \ref{sec:intlib} gives more information on the
interface between \textsc{Ninja} and the libraries of Master
Integrals.

Since this paper is rather technical, the reader which is mostly
interested in the usage of the library might want to read Sections
\ref{sec:usage} and \ref{sec:examples} first, referring to the
previous sections or the appendices at a later time if needed.

\section{Integrand Reduction via Laurent Expansion}
\label{sec:laurent}

In this section we review the \emph{Integrand Reduction via
  Laurent-expansion method}~\cite{Mastrolia:2012bu} for the
computation of one-loop integrals.  The method is based on the
systematic application of the Laurent series expansion on the
universal integrand decomposition of one-loop amplitudes, which allows
to reduce any amplitude as a linear combination of known Master
Integrals.

\subsection{Universal one-loop decomposition}
\label{sec:decomposition}

A generic contribution $\M$ to an $n$-point one-loop amplitude in
dimensional regularization has the form
\begin{equation}
  \label{eq:integral}
  \M = h(\mu_R^2,d) \int d^d \bar q \; \I = h(\mu_R^2,d) \int d^d \bar q\, \frac{\N(\bar q)}{D_0\cdots D_{n-1}}.
\end{equation}
In the previous equation, the integrand $\I$ is a \emph{rational
  function} of the components of the $d$-dimensional loop momentum
$\bar q$, with $d=4-2 \epsilon$.  The numerator $\N(\bar q)$ is a
process-dependent polynomial in $\bar q$, while the denominators $D_i$
are quadratic polynomials in $\bar q$ and correspond to Feynman loop
propagators,
\begin{equation}
  \label{eq:loopdenom}
  D_i = (\bar q + p_i)^2-m_i^2.
\end{equation}
The function $h$ appearing in Eq.~\eqref{eq:integral} is a
conventional normalization factor given by~\cite{Ellis:2007qk}
\begin{equation}
  \label{eq:hmurd}
  h(\mu_R^2,d) = h(\mu_R^2,4-2\epsilon) = \frac{\mu_R^{2\epsilon}}{i \pi^{2-\epsilon}} \frac{\Gamma(1-2\epsilon)}{\Gamma^2(1-\epsilon)\Gamma(1+\epsilon)},
\end{equation}
as a function of the renormalization scale $\mu_R^2$ and the dimension
$d=4-2\epsilon$.  The $d$-dimensional loop momentum $\bar q$ can be
split into a four-dimensional part $q$ and a
$(-2\epsilon)$-dimensional part $\vec{\mu}$ as
\begin{equation}
  \label{eq:qandmu}
  \bar q = q + \vec{\mu} \ ,  \qquad  \bar q^2 = q^2 -\mu^2.
\end{equation}
The numerator $\N$ will therefore be a polynomial in the four
components of $q$ and the extra-dimensional variable $\mu^2$.

Every one-loop integrand in dimensional regularization can be
decomposed as sum of integrands having five or less loop denominators
\cite{Ossola:2006us,Ellis:2007br}
\begin{equation}
  \label{eq:integranddec}
  \I \equiv \frac{\N}{D_0\cdots D_{n-1}} = \sum_{k=1}^5\sum_{\{i_1,\cdots, i_k\}}\frac{\Delta_{i_1\cdots i _k}}{D_{i_1}\cdots D_{i_k}},
\end{equation}
where the second sum on the r.h.s.\ runs over all the subsets of the
denominator indexes $\{0,\ldots, n-1\}$ containing $k$ elements.  The
\emph{residues} $\Delta_{i_1\cdots i_k}$ appearing in
Eq.~\eqref{eq:integranddec} are irreducible polynomials, i.e.\
polynomials which do not contain terms proportional to the
corresponding loop denominators $D_{i_1},\ldots,D_{i_k}$.  These
residues have a universal, process-independent parametric form in
terms of unknown, process-dependent coefficients.

For any set of denominators $D_{i_1},\ldots,D_{i_k}$ with $k\leq 5$,
one can build a $4$-dimensional basis of massless momenta
$\mathcal{E}^{(i_1\cdots
  i_k)}=\{e_1,e_2,e_3,e_4\}$~\cite{Forde:2007mi,delAguila:2004nf,Ossola:2006us,Mastrolia:2010nb}.
The first two elements of the basis are linear combinations of two
external momenta $K_1$, $K_2$ of the sub-diagram identified by the
considered set of loop denominators.  More explicitly, we define
\begin{equation}
  e_1^\mu = \frac{1}{1-r_1 r_2}\left(K_1^\mu - r_1\, K_2^\mu\right), \qquad e_2^\mu = \frac{1}{1-r_1 r_2}\left(K_2^\mu - r_2\, K_1^\mu\right),
\end{equation}
with
\begin{align*}
  & K_1^\mu = p_{i_1}^\mu-p_{i_k}^\mu, \qquad K_2^\mu = p_{i_2}^\mu-p_{i_1}^\mu, \qquad
  r_1 = \frac{K_1^2}{\gamma}, \qquad r_2 = \frac{K_2^2}{\gamma}, \nn
  & \gamma = (K_1\cdot K_2) \left(1 + \sqrt{1 - \frac{K_1^2 K_2^2}{(K_1\cdot K_2)^2}}\right),
\end{align*}
where the momenta $p_i$ were defined in Eq.~\eqref{eq:loopdenom}.  If
the sub-diagram has less than two independent external momenta, the
remaining ones are substituted by arbitrary reference vectors in the
definition of $e_1$ and $e_2$.  The momenta $e_3$ and $e_4$ are
instead chosen to be orthogonal to the first two and can be defined
using the spinor notation as
\begin{equation}
  e_3^\mu = \frac{\langle e_1\, \gamma^\mu\, e_2]}{2}, \qquad e_4^\mu = \frac{\langle e_2\, \gamma^\mu\, e_1]}{2}.
\end{equation}
They satisfy the property $(e_3\cdot e_4) = -(e_1\cdot e_2)$.  For
subsets of $k=4$ denominators, we also define the vectors $v$ and
$v_\perp$
\begin{equation}
  v^\mu = (e_4\cdot K_3)\, e_3^\mu + (e_3\cdot K_3)\, e_4^\mu, \qquad v_\perp^\mu = (e_4\cdot K_3)\, e_3^\mu - (e_3\cdot K_3)\, e_4^\mu,
\end{equation}
with $K_3^\mu = p_{i_3}^\mu - p_{i_2}^\mu$.  We observe that the
vector $v_\perp$ is orthogonal to all the external legs of the
sub-diagram identified by the four denominators.

By expanding the four dimensional part $q$ of the loop momentum $\bar
q$ in the basis $\mathcal{E}^{(i_1\cdots i_k)}$, the numerator and the
denominators can be written as polynomials in the coordinates
$\mathbf{z} \equiv (z_1,z_2,z_3,z_4,z_5) = (x_1,x_2,x_3,x_4,\mu^2)$,
\begin{equation}
  \N(\bar q) = \N(q,\mu^2) = \N(x_1,x_2,x_3,x_4,\mu^2) = \N(\mathbf{z}),
\end{equation}
with
\begin{equation}
  q^\nu  = -p^\nu_{i_1} + \frac{1}{e_1\cdot e_2} \left(  x_1 e_1^\nu + x_2 e^\nu_2 - x_3 e^\nu_3 - x_4 \ e^\nu_4 \right).
\end{equation}
The coordinates $x_i$ can be written as scalar products
\begin{equation}
  x_1 = (l_{i_1}\cdot e_2), \quad
  x_2 = (l_{i_1}\cdot e_1), \quad
  x_3 = (l_{i_1}\cdot e_4),\quad
  x_4  = (l_{i_1}\cdot e_3),
\end{equation}
where $l_{i_1}\equiv (q+p_{i_1})$.  For $k=4$ we also consider the
alternative expansion of the loop momentum
\begin{equation}
  q^\nu  = -p^\nu_{i_1} + \frac{1}{e_1\cdot e_2} \left(  x_1 e_1^\nu + x_2 e^\nu_2 \right) + \frac{1}{v^2} \left( x_{3,v}\, v^\nu_3 - x_{4,v} \ v^\nu_4 \right).
\end{equation}
with
\begin{equation}
  x_{3,v} = (l_{i_1}\cdot v),\quad
  x_{4,v}  = (l_{i_1}\cdot v_\perp).
\end{equation}

The universal parametric form of the residues $\Delta_{i_1\cdots i_k}$
in a renormalizable theory is
\cite{Ossola:2006us,Ellis:2007br,Mastrolia:2012an}
\begin{align}
  \Delta_{i_1 i_2 i_3 i_4 i_5} ={}& c_0\, \mu^2 \nn
  \Delta_{i_1 i_2 i_3 i_4} ={}& c_0 + c_1 x_{4,v} + \mu^2 \left ( c_2 + c_3 x_{4,v} + \mu^2 c_4 \right )\nn
  {\Delta}_{i_1 i_2 i_3} ={}& c_0 + c_1 x_4 + c_2 x_4^2 + c_3 x_4^3
  + c_4 x_3 + c_5 x_3^2 + c_6 x_3^3
+ \mu^2 \left (c_7 + c_8 x_4 + c_9 x_3 \right ) \nn
{\Delta}_{i_1 i_2} ={}& c_0  + c_1 x_1  + c_2 x_2^2  + c_3 x_4 + c_4 x_4^2 + c_5 x_3
+   c_6 x_3^2 + c_7 x_1 x_4
+ c_8 x_1 x_3  + c_9 \mu^2  \nn
{\Delta}_{i_1} ={}& c_0  + c_1 x_2 + c_2 x_1 + c_3 x_4
+   c_4 x_3.
\label{eq:parametricresidues}
\end{align}
where we understand that the unknown coefficients $c_j$ depend on the
indexes of the residue (e.g.\ $c_j=c_j^{(i_1\cdots i_k)}$), while the
scalar products $x_i$ and $x_{i,v}$ depend on both the indexes of the
residue and the loop momentum $q$.  The parametrization in
Eq.~\eqref{eq:parametricresidues} can be extended to effective and
non-renormalizable theories where the rank of the numerator can be
larger than the number of loop propagators \cite{Mastrolia:2012bu}.
More details on the higher-rank case are given in
Appendix~\ref{sec:hr}.

Most of the terms appearing in Eq.~\eqref{eq:parametricresidues} are
\emph{spurious}, i.e.\ they vanish after integration and do not
contribute to the final result.  The amplitude $\M$ can thus be
expressed as a linear combination of Master Integrals, corresponding
to the non-spurious terms of the integrand decomposition, namely
\begin{align}
  \M = {}&
\sum_{\{i_1, i_2, i_3, i_4\}}\bigg\{
          c_{0}^{ (i_1 i_2 i_3 i_4)} I_{i_1 i_2 i_3 i_4} +  
          c_{4}^{ (i_1 i_2 i_3 i_4)} I_{i_1 i_2 i_3 i_4}[\mu^4] 
\bigg\} \pagebreak[1] \nn
     & +
\sum_{\{i_1, i_2, i_3\}}\bigg\{
          c_{0}^{ (i_1 i_2 i_3)} I_{i_1 i_2 i_3} +
          c_{7}^{ (i_1 i_2 i_3)} I_{i_1 i_2 i_3}[\mu^2]
\bigg\} \pagebreak[1] \nn
     & +
\sum_{\{i_1, i_2\}}\bigg\{
          c_{0}^{ (i_1 i_2)} I_{i_1 i_2} 
        + c_{1}^{ (i_1 i_2)} I_{i_1 i_2}[(q+p_{i_1})\cdot e_2 ] \nn
        & \qquad
        + c_{2}^{ (i_1 i_2)} I_{i_1 i_2}[((q+p_{i_1})\cdot e_2)^2 ]         +  c_{9}^{ (i_1 i_2)} I_{i_1 i_2}[\mu^2]  \bigg\} \pagebreak[1] \nn 
& + 
\sum_{i_1}
      c_{0}^{ (i_1)} I_{i_1},   \label{eq:integraldecomposition}
\end{align}
where
\begin{equation*}
  I_{i_1 \cdots i_k}[\alpha] \equiv h(\mu_R^2,d)\, \int d^d \bar q {\alpha \over D_{i_1} \cdots D_{i_k} }, 
\qquad
I_{i_1 \cdots i_k} \equiv   I_{i_1 \cdots i_k}[1].
\end{equation*}
The coefficients of this linear combination can be identified with a
subset of the coefficients of the parametric residues in
Eq.~\eqref{eq:parametricresidues}.  Since all the Master Integrals
appearing in Eq.~\eqref{eq:integraldecomposition} are known, the
problem of the computation of an arbitrary one-loop amplitude can be
reduced to the problem of the determination of the non-spurious
coefficients appearing in the parametrization of the residues
$\Delta_{i_1\cdots i_k}$.

\subsection{The Laurent expansion method}
The coefficients appearing in the integrand decomposition can be
computed by evaluating the integrand on \emph{multiple cuts}, i.e.\ on
values of the loop momentum $\bar q$ such that a subset of loop
denominators vanish \cite{Ossola:2007ax}.  More in detail, the
coefficients of a $k$-point residue $\Delta_{i_1\cdots i_k}$ can be
determined by evaluating the integrand on the corresponding $k$-ple
cut $D_{i_1}=\cdots =D_{i_k}=0$.  For these values of the loop
momentum, the only non-vanishing contributions of the integrand
decomposition are the ones coming from the residue in consideration
and from all the higher-point residues which have $\{D_{i_1},\ldots
,D_{i_k}\}$ as a subset of their loop denominators.

Within the original integrand reduction
algorithm~\cite{Ossola:2007ax,Mastrolia:2008jb,Mastrolia:2010nb}, the
coefficients are computed by sampling the numerator of the integrand
on a finite subset of the on-shell solutions, subtracting all the
non-vanishing contributions coming from higher-point residues, and
finally solving the resulting linear system of equations.  This is
therefore a top-down approach, where higher-point residues are
computed first, starting from $k=5$, and systematically subtracted
from the integrand for the evaluation of lower-point contributions.
These are referred to as subtractions at the integrand level.

The integrand reduction via Laurent expansion method, presented in
Ref.~\cite{Mastrolia:2012bu}, improves this reduction strategy by
elaborating on techniques proposed
in~\cite{Forde:2007mi,Badger:2008cm}.  Whenever the analytic
dependence of the integrand on the loop momentum is known, this
approach allows to compute the coefficients of a residue
$\Delta_{i_1\cdots i_k}$ by performing a Laurent expansion of the
integrand with respect to one of the components of the loop momentum
which are not fixed by the on-shell conditions of the corresponding
multiple cut $D_{i_1}=\cdots = D_{i_k}=0$.  In the asymptotic limit defined by this Laurent expansion,
both the integrand and the higher-point subtractions exhibit the same
polynomial behavior as the residue.  Therefore one can directly
identify the unknown coefficients with the ones of the Laurent
expansion of the integrand, corrected by the contributions coming from
higher-point residues.

Hence, by choosing a suitable Laurent expansion, one obtains a
diagonal system of equations for the coefficients of the residues,
while the subtractions of higher-point contributions can be
implemented as \emph{corrections at the coefficient level} which
replace the subtractions at the integrand level of the original
algorithm.  Since the polynomial structure of the residues is
universal and does not depend on the process, the parametric form of
the coefficient-level corrections can be computed once and for all, in
terms of a subset of the higher-point coefficients.  More in detail,
the corrections at the coefficient level are known functions of a
subset of the coefficients of 3- and 2-point residues.  In particular,
no subtraction term coming from 4- and 5-point contributions is ever
needed.  This allows to skip the computation of the (spurious) 5-point
contributions entirely, and to completely disentangle the
determination of 4-point residues from the one of lower point
contributions.

In the following, we address more in detail the computation of 5-, 4-,
3-, 2-, and 1-point residues, also commonly known as \emph{pentagons},
\emph{boxes}, \emph{triangles}, \emph{bubbles} and \emph{tadpoles}
respectively.  For simplicity, we focus on renormalizable theories,
where (up to a suitable choice of gauge) the maximum allowed rank of
the integrand is equal to the number of loop denominators and the most
general parametrization of the residues is the one given in
Eq.~\eqref{eq:parametricresidues}.  \textsc{Ninja} can also be used
for the computation of integrals whose rank exceeds the number of
denominators by one.  The extension of the method to the higher-rank
case is discussed in Appendix~\ref{sec:hr}.

\paragraph{5-point residues}
As mentioned above, pentagon contributions are spurious.  Within the
original integrand reduction algorithm, their computation is needed
because they appear in the subtractions at the integrand level
required for the evaluation of lower-point contributions.  A 5-point
residue only has one coefficient, which can easily be computed by
evaluating the numerator of the integrand on the corresponding 5-ple
cut.  Within the Laurent expansion approach, the subtraction terms
coming from five-point residues always vanish in the asymptotic limits
we consider, therefore their computation can be skipped.  For this
reason, in the library \textsc{Ninja} the computation of pentagons is
disabled by default, even though it can be enabled for debugging
purposes, as explained in Section \ref{sec:runtime}.

\paragraph{4-point residues}
The coefficient $c_0$ of a box contribution $\Delta_{i_1 i_2 i_3 i_4}$
can be determined via four-dimensional quadruple cuts
\cite{Britto:2004nc}.  A quadruple cut $D_{i_1}=\cdots=D_{i_4}=0$ in
four dimensions (i.e.\ $\bar q = q$, $\mu^2=0$) has two solutions,
$q_+$ and $q_-$.  The coefficient $c_0$ can be expressed in terms of
these solutions as
\begin{equation}
  \label{eq:c4}
 c_0=\frac{1}{2} \left(\left. \frac{\N(q)}{\prod_{j\neq i_1,i_2,i_3,i_4}D_j}\right|_{q=q_+}+\left.\frac{\N(q)}{\prod_{j\neq i_1,i_2,i_3,i_4}D_j}\right|_{q=q_-}\right).
\end{equation}
Given the simplicity of Eq.~\eqref{eq:c4}, this is the only
coefficient which \textsc{Ninja} computes in the same way as the
traditional algorithm.  The coefficient $c_4$ can instead be
determined by evaluating the integrand on $d$-dimensional quadruple
cuts in the asymptotic limit of large $\mu^2$~\cite{Badger:2008cm}.  A
$d$-dimensional quadruple cut has an infinite number of solutions
which can be parametrized by the ($-2\epsilon$)-dimensional variable
$\mu^2$.  These solutions become simpler in the considered limit,
namely
\begin{equation}
  \label{eq:qmuexp}
  q_{\pm}^\nu = -p_{i_1}^\nu + a^\nu \pm \sqrt{\mu^2+\beta}\, v_\perp^\nu\; \overset{\mu^2\rightarrow\infty}{=}\; \pm \sqrt{\mu^2} \, v_\perp^\nu + \O(1),
\end{equation}
where the vector $a^\nu$ and the constant $\beta$ are fixed by the cut
conditions. The coefficient $c_4$ is non-vanishing only if the rank of
the numerator is greater or equal to the number of loop denominators.
In a renormalizable theory, it can be found in the
$\mu^2\rightarrow\infty$ asymptotic limit as the leading term of the
Laurent expansion of the integrand
\begin{equation}
   \left. \frac{\N(q,\mu^2)}{\prod_{j\neq i_1,i_2,i_3,i_4}D_j} \right|_{q=\sqrt{\mu^2} v_\perp+\O(1)} = c_4\, \mu^4 +\O(\mu^3).
\end{equation}
The other coefficients of the boxes are spurious and, since they
neither contribute to the final result nor to the subtraction terms,
their computation can be skipped.

\paragraph{3-point residues}
The coefficients of the residues of a generic triangle contribution
$\Delta_{i_1 i_2 i_3}$ can be determined by evaluating the integrand
on the solutions of the corresponding $d$-dimensional triple
cut~\cite{Forde:2007mi}.  These can be parametrized by the variable
$\mu^2$ and the free parameter $t$,
\begin{equation}
  \label{eq:triplecutsolutions}
  q_+^\nu = -p_{i_1}^\nu +  a^\nu+ t\, e_3^\nu + \frac{\beta + \mu^2}{2\, t\, (e_3\cdot e_4)}\, e_4^\nu, \qquad   q_- = -p_{i_1}^\nu + a^\nu+ \frac{\beta + \mu^2}{2\, t\, (e_3\cdot e_4)}\, e_3^\nu+t\, e_4^\nu,
\end{equation}
where the vector $a^\nu$ and the constant $\beta$ are fixed by the cut
conditions $D_{i_1}=D_{i_2}=D_{i_3}=0$.  The momentum $a^\nu$ is a
linear combination of $e_1$ and $e_2$ and is therefore orthogonal to
$e_3$ and $e_4$.  On these solutions, the non-vanishing contributions
to the integrand decomposition are the ones of the residue
$\Delta_{i_1 i_2 i_3}$, as well as the ones of the boxes and pentagons
which share the three cut denominators.  However, after performing a
Laurent expansion for large $t$ and dropping the terms which vanish in
this limit, the pentagon contributions vanish, while the box
contributions are constant in $t$ but they also vanish when taking the
average between the parametrizations $q_+$ and $q_-$ of
Eq.~\eqref{eq:triplecutsolutions}.  More explicitly,
\begin{alignat}{3} \frac{\N(q_\pm,\mu^2)}{\prod_{j\neq i_1,i_2,i_3}D_j} & = \Delta_{i_1 i_2 i_3} + {} & &{}\sum_{j}\frac{\Delta_{i_1 i_2 i_3 j}}{D_j}& & + \sum_{jk}\frac{\Delta_{i_1 i_2 i_3 jk}}{D_jD_k} \nn
  &= \Delta_{i_1 i_2 i_3} + {} & & {} d_1^\pm + d_2^\pm\, \mu^2 & &+ \O(1/t),\quad \qquad \textrm{with }d_i^+ + d_i^- = 0.
\label{eq:triangledecexp}
\end{alignat}
Moreover, the expansion of the integrand is given by
\begin{align}
           \frac{\N(q_+,\mu^2)}{\prod_{j\neq i_1,i_2,i_3}D_j}  ={}& n_0^++n_7^+\, \mu^2+(n_4+n_9\, \mu^2)\, {t} +n_5\, {t^2} +n_6\, {t^3} +\O(1/{t}), \nn
           \frac{\N(q_-,\mu^2)}{\prod_{j\neq i_1,i_2,i_3}D_j} ={}& n_0^-+n_7^-\, \mu^2+(n_1+n_8\, \mu^2)\, {t} +n_2\, {t^2} +n_3\, {t^3} +\O(1/{t}) \label{eq:trianglenumexp}
\end{align}
and it has the same polynomial behavior as the expansion of the
residue $\Delta_{i_1 i_2 i_3}$,
\begin{align}
        \Delta_{i_1 i_2 i_3}(q_+,\mu^2) ={}& c_0 + c_7\, \mu^2 + (c_4+c_9\, \mu^2)\, (e_3\cdot e_4)\, {t} + c_5\, (e_3\cdot e_4)^2\, {t^2} + c_6\, (e_3\cdot e_4)^3\, {t^3} +\O(1/{t}), \nn
       \Delta_{i_1 i_2 i_3}(q_-,\mu^2) ={}& c_0 + c_7\, \mu^2 + (c_1+c_8\, \mu^2)\, (e_3\cdot e_4)\, {t} + c_2\, (e_3\cdot e_4)^2\, {t^2} + c_3\, (e_3\cdot e_4)^3\, {t^3} +\O(1/{t}) \label{eq:triangledeltaexp}.
\end{align}
By comparison of Eq.~\eqref{eq:triangledecexp},
\eqref{eq:trianglenumexp} and \eqref{eq:triangledeltaexp} one can
directly identify the ten triangle coefficients as the corresponding
terms of the expansion of the integrand,
\begin{equation}
  c_{0,7}=\frac{1}{2}(n_{0,7}^+ + n_{0,7}^-), \qquad c_{1,4,8,9} = \frac{n_{1,4,8,9}}{(e_3\cdot e_4)},  \qquad c_{2,5} = \frac{n_{2,5}}{(e_3\cdot e_4)^2}, \qquad c_{3,6} = \frac{n_{3,6}}{(e_3\cdot e_4)^3}.
\end{equation}
Hence, with the Laurent expansion method, the determination of the
3-point residues does not require any subtraction of higher-point
terms.

\paragraph{2-point residues}
The coefficients of a generic 2-point residue $\Delta_{i_1 i_2}$ can
be evaluated on the on-shell solutions of the corresponding double cut
$D_{i_1}=D_{i_2}=0$, which can be parametrized as
\begin{align}
  \label{eq:doublecutsolutions}
  q_+ & = -p_0 + x\, e_1^\nu + (\alpha_0+x\, \alpha_1)\, e_2^\nu + t\, e_3^\nu + \frac{\beta_0 + \beta_1\, x + \beta_2\, x^2 + \mu^2}{2\, t\, (e_3\cdot e_4)}\, e_4^\nu, \nn
  q_- & = -p_0 + x\, e_1^\nu + (\alpha_0+x\, \alpha_1)\, e_2^\nu + \frac{\beta_0 + \beta_1\, x + \beta_2\, x^2 + \mu^2}{2\, t\, (e_3\cdot e_4)}\, e_3^\nu+t\, e_4^\nu,
\end{align}
in terms of the three free parameters $x$, $t$ and $\mu^2$, while the
constants $\alpha_i$ and $\beta_i$ are fixed by the on-shell
conditions.  After evaluating the integrand on these solutions and
performing a Laurent expansion for $t\rightarrow\infty$, the only
non-vanishing subtraction terms come from the triangles,
\begin{align}
  \frac{\N(q_\pm,\mu^2)}{\prod_{j\neq i_1,i_2}D_j} &
  = \Delta_{i_1 i_2} + \sum_{j}\frac{\Delta_{i_1 i_2 j}}{D_j}
  +\sum_{jk}\frac{\Delta_{i_1 i_2 jk}}{D_jD_k} +
  \sum_{jkl}\frac{\Delta_{i_1 i_2 jkl}}{D_jD_kD_l} \nn
  & = \Delta_{i_1 i_2}+\sum_{j}\frac{\Delta_{i_1 i_2 j}}{D_j}+\O(1/t).   \label{eq:bubbledecexp}
\end{align}
Even though the integrand and the subtraction terms are rational
functions, in the asymptotic limit they both have the same polynomial
behavior as the residue, namely
\begin{align}
        \frac{\N(q_+,\mu^2)}{\prod_{j\neq i_1,i_2}D_j} ={}& n_0+n_9\, {\mu^2}+n_1\, {x} +n_2\, {x^2} - \big(n_5 + n_8 {x}\big) {t} +n_6\, {t^2} +\O(1/{t}) \nn
        \frac{\N(q_-,\mu^2)}{\prod_{j\neq i_1,i_2}D_j} ={}& n_0+n_9\, {\mu^2}+n_1\, {x} +n_2\, {x^2} - \big(n_3 + n_7 {x}\big) {t} +n_4\, {t^2} +\O(1/{t}) \label{eq:bubblenumexp} \pagebreak[1]  \\
   \frac{\Delta_{i_1 i_2 j}(q_+,\mu^2)}{D_j} ={}&  c_{s_3,0}^{(j)}+ c_{s_3,9}^{(j)}\, {\mu^2}+ c_{s_3,1}^{(j)}\, {x} + c_{s_3,2}^{(j)}\, {x^2} - \Big( c_{s_3,5}^{(j)} +  c_{s_3,8}^{(j)} {x}\Big) {t} + c_{s_3,6}^{(j)}\, {t^2}+\O(1/{t}) \nn
   \frac{\Delta_{i_1 i_2 j}(q_-,\mu^2)}{D_j} ={}&  c_{s_3,0}^{(j)}+ c_{s_3,9}^{(j)}\, {\mu^2}+ c_{s_3,1}^{(j)}\, {x} + c_{s_3,2}^{(j)}\, {x^2} - \Big( c_{s_3,3}^{(j)} +  c_{s_3,7}^{(j)} {x}\Big) {t} + c_{s_3,4}^{(j)}\, {t^2}+\O(1/{t}) \label{eq:bubblessubexp} \pagebreak[1] \\
   \Delta_{i_1 i_2}(q_+,\mu^2) ={}& c_0+c_9\, {\mu^2}+c_1\, (e_1\cdot e_2)\, {x} +c_2 \, (e_1\cdot e_2)^2\,  {x^2}\nn
   & + \Big(c_5 + c_8 \, (e_1\cdot e_2)\,  {x}\Big)\, (e_3\cdot e_4)\, {t} +c_6\, (e_3\cdot e_4)^2\, {t^2}+\O(1/{t}) \nn
   \Delta_{i_1 i_2}(q_-,\mu^2) ={}& c_0+c_9\, {\mu^2}+c_1\, (e_1\cdot e_2)\, {x} +c_2\, (e_1\cdot e_2)^2\, {x^2} \nn
   & +\Big(c_3 + c_7 \, (e_1\cdot e_2)\, {x}\Big)\, (e_3\cdot e_4)\, {t} +c_4\, (e_3\cdot e_4)^2\, {t^2}+\O(1/{t}). \label{eq:bubbledeltaexp}
\end{align}
The coefficients $c_{s_3,i}^{(j)}$ of the expansion of the
subtractions terms in Eq.s \eqref{eq:bubblessubexp} are known
parametric functions of the triangle coefficients.  Hence, the
subtraction of the triangle contributions can be implemented by
applying coefficient-level corrections to the terms appearing in the
expansion of the integrand.  More explicitly, by inserting
Eq.s~\eqref{eq:bubblenumexp}, \eqref{eq:bubblessubexp} and
\eqref{eq:bubbledeltaexp} in Eq.~\eqref{eq:bubbledecexp} one gets
\begin{align}
  c_{0,9} & = n_{0,9} - \sum_j c_{s_3,0,9}^{(j)}, \nn
  c_{1,3,5}&  = \frac{1}{(e_1\cdot e_2)}\Big( n_{1,3,5} - \sum_j c_{s_3,1,3,5}^{(j)}\Big), \nn
  c_{2,4,6,7,8} & = \frac{1}{(e_1\cdot e_2)^2}\Big( n_{2,4,6,7,8} - \sum_j c_{s_3,2,4,6,7,8}^{(j)}\Big).
  \label{eq:bubcoeffs}
\end{align}

\paragraph{1-point residues}
The only non-spurious coefficient $c_0$ of a tadpole residue
$\Delta_{i_1}$ can be computed by evaluating the integrand on
solutions of the single cut $D_{i_1}=0$.  For this purpose, one can
consider 4-dimensional solutions of the form
\begin{equation}
  q^\nu_+ = -p_{i_1}^\nu + t\, e_3^\nu + \frac{m_{i_1}^2}{2\, t\, (e_3\cdot e_4)}\, e_4^\nu,
\end{equation}
parametrized by the free variable $t$.  In the asymptotic limit
$t\rightarrow \infty$, only bubble and triangle subtraction terms are
non-vanishing,
\begin{align} \frac{\N(q_+)}{\prod_{j\neq i_1}D_j} & = \Delta_{i_1} 
+ \sum_{j}\frac{\Delta_{i_1  j}}{D_j}+ \sum_{jk}\frac{\Delta_{i_1 jk}}{D_j D_k}+ \sum_{jkl}\frac{\Delta_{i_1 jkl}}{D_jD_k D_l} \nn
&= \Delta_{i_1}+ \sum_{j}\frac{\Delta_{i_1  j}}{D_j}+ \sum_{jk}\frac{\Delta_{i_1 jk}}{D_j D_k} + \O(1/t).
\label{eq:tadpoledecexp}
\end{align}
Similarly to the case of the 2-point residues, in this limit the
integrand and the subtraction terms exhibit the same polynomial
behavior as the residue, i.e.\
\begin{align}
  \frac{\N(q_+)}{\prod_{j\neq i_1}D_j} & = n_0 + n_4\, (e_3\cdot e_4) \, t + \O(1/t) \\
  \frac{\Delta_{i_1 j}(q_+)}{D_j} & = c_{s_2,0}^{(j)} + c_{s_2,4}^{(j)}\, (e_3\cdot e_4) \, t + \O(1/t) \\
  \frac{\Delta_{i_1 j k}(q_+)}{D_j D_k} & = c_{s_3,0}^{(jk)} + c_{s_3,4}^{(jk)}\, (e_3\cdot e_4) \, t + \O(1/t) \\
  \Delta_{i_1}(q_+) & = c^{(i_1)}_0 + c^{(i_1)}_4\, t + \O(1/t).
\end{align}
Putting everything together, the coefficient of the tadpole integral
can be identified with the corresponding one in the expansion of the
integrand, corrected by coefficient-level subtractions from bubbles
and triangles
\begin{equation}
  c_0 = n_0 - \sum_j c_{s_2,0}^{(j)}  - \sum_{jk} c_{s_3,0}^{(jk)}.
  \label{eq:tadcoeffs}
\end{equation}
The subtraction terms $c_{s_2,0}^{(j)}$ and $c_{s_3,0}^{(jk)}$, coming
from 2-point and 3-point contributions respectively, are known
parametric functions of the coefficients of the corresponding
higher-point residues.

\section{Semi-numerical implementation}
\label{sec:ninja}
The {\sc C++} library {\sc Ninja} provides a semi-numerical
implementation of the Laurent expansion method described in
Section~\ref{sec:laurent}.  The Laurent series expansion is typically
an analytic operation, but since a one-loop integrand is a rational
function of the loop variables, its expansion can be obtained via a
partial fraction decomposition between the numerator and the
denominators.  This is implemented in \textsc{Ninja} via a simplified
polynomial-division algorithm, which takes as input the coefficients
of a parametric expansion of the numerator $\N$ and computes the
leading terms of the quotient of the polynomial division with respect
to the uncut denominators.  In this section we describe the input
needed for the reduction performed by \textsc{Ninja} and we give
further details about the implementation of the reduction.

All the types, classes and functions provided by the \textsc{Ninja}
library are defined inside the \texttt{ninja} namespace.  In
particular, the types \texttt{Real} and \texttt{Complex} are aliases
for \texttt{double} and \texttt{std::complex<double>}, unless the
library was compiled in quadruple precision.  Classes for real and
complex momenta are defined as \texttt{RealMomentum} and
\texttt{ComplexMomentum} respectively.  They are wrappers of
four-dimensional arrays of the corresponding floating-point types,
which overload arithmetic and subscript operators.  More in detail, an
instance \texttt{p} of one of these classes represents a momentum $p$
according to the representation
\begin{equation*}
  \texttt{p} = \{\texttt{p[0]},\texttt{p[1]},\texttt{p[2]},\texttt{p[3]}\} = \{E_p,x_p,y_p,z_p\},
\end{equation*}
i.e.\ with the energy in the zeroth component, followed by the spatial
components.

\subsection{Input}
\label{sec:input}
The inputs needed from the reduction algorithm implemented in {\sc
  Ninja} are the momenta $p_i$ and the masses $m_i$ of the loop
denominators defined in Eq.~\eqref{eq:loopdenom}, besides the
numerator $\N(q,\mu^2)$ of the integrand.  The latter must be cast in
four different forms, one of which is optional.  The \textsc{C++}
implementation requires the numerator to be an instance of a class
inherited from the abstract class \texttt{ninja::Numerator}.  The
latter defined as

{\small \begin{lstlisting}
class Numerator {

public:

  virtual Complex evaluate(const ninja::ComplexMomentum & $\(q\)$,
                           const ninja::Complex & $\(\mu^{2} \)$,
                           int cut,
                           const ninja::PartitionInt partition[])
                           = 0;

  virtual void muExpansion(const ninja::ComplexMomentum $\(v\)$[],
                           const ninja::PartitionInt partition[],
                           ninja::Complex c[]) {}

  virtual void t3Expansion(const ninja::ComplexMomentum & $\(v_{0}\)$,
                           const ninja::ComplexMomentum & $\(v_{3}\)$,
                           const ninja::ComplexMomentum & $\(v_{4}\)$,
                           const ninja::Complex & $\(\beta\)$,
                           int mindeg,
                           int cut,
                           const ninja::PartitionInt partition[],
                           ninja::Complex c[]) = 0;

  virtual void t2Expansion(const ninja::ComplexMomentum & $\(v_{1}\)$,
                           const ninja::ComplexMomentum & $\(v_{2}\)$,
                           const ninja::ComplexMomentum & $\(v_{3}\)$,
                           const ninja::ComplexMomentum & $\(v_{4}\)$,
                           const ninja::Complex $\(\beta\)$[],
                           int mindeg,
                           int cut,
                           const ninja::PartitionInt partition[],
                           ninja::Complex c[]) = 0;

  virtual ~Numerator() {}

};
\end{lstlisting}}

The input parameters \texttt{cut} and \texttt{partition} are common to
more methods and give information about the multiple cut where
\textsc{Ninja} is currently evaluating the numerator.  Although this
information is not always necessary, there might be occasions where it
could be useful for an efficient evaluation of the numerator.  The
integer \texttt{cut} is equal to $k$ if the numerator is being
evaluated on a $k$-ple cut, with $k=1,2,3,4$.  This parameter is not
given in the method \texttt{muExpansion} because the latter is always
evaluated on quadruple cuts (i.e.\ $\texttt{cut}=4$).  The parameter
\texttt{partition} points to an array of integers (namely of integer
type \texttt{ninja::PartitionInt}), with length equal to \texttt{cut},
containing the indexes of the cut numerators.  If the user asks to
perform a global test (see Section~\ref{sec:runtime}), the numerator
will also be evaluated outside the solutions of the multiple cuts, in
which case the parameter \texttt{cut} will be set to zero.  As an
example, if the method \texttt{t3Expansion} is evaluated on the 3-ple
cut $D_0=D_2=D_5=0$ for the determination of the 3-point residue
$\Delta_{025}$, we will have $\texttt{cut}=3$,
$\texttt{partition[0]}=0$, $\texttt{partition[1]}=2$, and
$\texttt{partition[2]}=5$.  The returned expansion only needs to be
valid for the cut specified (e.g.\ $D_0=D_2=D_5=0$ in the previous
example).

Here is a detailed description of each method of the class
\texttt{ninja::Numerator}, for a generic numerator of an $n$-point
integrand of rank $r$.  If the analytic expression of the integrand is
available, all these methods can be easily generated with the help of
the simple \textsc{Python} package \textsc{NinjaNumGen}, which is
distributed with the library and whose usage is described in
Section~\ref{sec:integrand} and in Appendix~\ref{sec:ninjanumgen}.
The details which follow are only needed to those who prefer to
provide an alternative implementation of the required methods without
\textsc{NinjaNumGen}.

\subsubsection*{The method \texttt{Numerator::evaluate($q$,$\mu^2$,cut,partition)}}
  It must return the value of the numerator $\N(q,\mu^2)$ evaluated at
  the (complex) values of $q$ and $\mu^2$ given as input.

  \subsubsection*{The method \texttt{Numerator::muExpansion($v$,partition,c)}}
  This is used for the computation of the Laurent expansion in $\mu^2$
  required to obtain the coefficient $c_4$ of the boxes.  In the
  renormalizable case, this method should compute the leading term of
  a parametric expansion in $t$ of the integrand defined by
  \begin{equation}
    q^\nu \rightarrow t\, v_\perp^\nu ,\qquad \mu^2 \rightarrow t^2\, v_\perp^2
  \end{equation}
  where $v_\perp$ is given by $v$\texttt{[0]}, i.e.\ by the zeroth
  entry of the array of momenta $v$.  For renormalizable theories this
  array will therefore only contain at most one element.  The method
  should write the leading coefficient of the expansion in the zeroth
  entry of the array pointed by the parameter \texttt{c}, i.e.\
\begin{lstlisting}
    c[0] = $\(\N\)$[$\(t^{r}\)$];
\end{lstlisting}
The generalization of this method to the higher-rank case is described
in Appendix~\ref{sec:hr}.  The implementation of this method is only
required when $r\geq n$.  It is also not needed if the user chooses to
disable the $\mu^2$-expansion method for the boxes, but in that case
more evaluations of the numerator will be needed and the computation
of the pentagons will not be skipped.

  \subsubsection*{The method
    \texttt{Numerator::t3Expansion($v_0$,$v_3$,$v_4$,$\beta$,mindeg,partition,c)}}
  This method is used for the computation of the coefficients of the
  residues of both the triangles and the tadpoles.  It must
  compute the coefficients of the terms $t^j\mu^{2k}$ for
  $j\in\{r,r-1,r-2,\ldots,r-\texttt{mindeg}\}$, given by substituting into
  the numerator the parametric expansion of the loop momentum defined
  by
  \begin{equation}
    \label{eq:ninjat3exp}
    q^\nu\rightarrow v_0^\nu + t\, v_3^\nu + \frac{\beta+\mu^2}{t}\, v_4^\nu, \qquad  v_3^2=v_4^2=0, \quad (v_3 \cdot v_4) = \frac{1}{2},
  \end{equation}
  as a function of the momenta $v_i^\nu$ and the constant $\beta$
  which are passed as parameters.  The maximum value of the parameter
  \texttt{mindeg} is $r-n+3$.  Since in a renormalizable theory $r \leq
  n$, and by definition of rank we have $j+2k\leq r$, in this case at
  most 6 terms can be non-vanishing in the specified range of $j$.
  The method should write these terms in the entries of the array
  pointed by \texttt{c}, ordered by decreasing powers of $t$.  Terms
  with the same power of $t$ should be ordered by increasing powers of
  $\mu^2$.  A pseudo-implementation will therefore look like
  \begin{lstlisting}
    int idx = 0;
    for (int j=$\(r\)$; j>=$\(r\)$-mindeg; --j)
      for (int k=0; 2*k<=$\(r\)$-j; ++k)
        c[idx++] = $\(\N\)$[$\(t^{j}\mu^{2k}\)$];
\end{lstlisting}  
\subsubsection*{The method \texttt{Numerator::t2Expansion($v_1$,$v_2$,$v_3$,$v_4$,$\beta$,mindeg,partition,c)}}
  In the current version of \textsc{Ninja}, this method is called
  during the computation of the coefficients of the bubbles.  It must compute the coefficients of the terms $t^jx^l\mu^{2k}$ for
  $j\in\{r,r-1,\ldots,r-\texttt{mindeg}\}$, given by the expansion
\begin{align}
  & q^\nu\rightarrow v_1^\nu + x\, v_2^\nu + t\, v_3^\nu + \frac{\beta_0+\beta_1\, x+\beta_2\, x^2+\mu^2}{t}\, v_4^\nu, \nn
  & v_2\cdot v_3=v_2\cdot v_4=v_3^2=v_4^2=0, \qquad (v_3\cdot v_4) = \frac{1}{2}
  \label{eq:ninjat2exp}
\end{align}
as a function of the momenta $v_i^\nu$ and the constants
$\beta_i\equiv\beta\texttt{[i]}$, which are passed as parameters to
the method.  The maximum value of \texttt{mindeg} is $r-n+2$.  In a
renormalizable theory, this implies that one can have at most 7
non-vanishing terms in this range of $j$.  It is worth observing that
the expansion in Eq.~\eqref{eq:ninjat2exp} can be obtained from the
previous one in Eq.~\eqref{eq:ninjat3exp} with the substitutions
\begin{equation*}
  v_0^\nu \rightarrow v_1^\nu + x\, v_2^\nu, \qquad \beta\rightarrow \beta_0+\beta_1\, x+\beta_2\, x^2, \qquad v_2\cdot v_3=v_2\cdot v_4=0.
\end{equation*}
The terms of the expansion must be stored in the entries of the array
pointed by \texttt{c}, ordered by decreasing powers of $t$.  Terms
with the same power of $t$ should be ordered from the lowest to the
highest with respect to the lexicographical order in the variables
$(x,\mu^2)$.  A pseudo-implementation will have the form
  \begin{lstlisting}
  int idx = 0;
  for (int j=$\(r\)$; j>=$\(r\)$-mindeg; --j)
    for (int l=0; l<=$\(r\)$-j; ++l)
      for (int k=0; 2*k<=$\(r\)$-j-l; ++k)
        c[idx++] = $\(\N\)$[$\(t^{j}x^{l}\mu^{2k}\)$];
\end{lstlisting}  

\subsection{Reduction via polynomial division}
\label{sec:polydiv}
For every phase-space point, {\sc Ninja} at run-time computes the
parametric solutions of the multiple cuts corresponding to each
residue.  The Laurent expansion of the integrand on these solutions is
performed via a simplified polynomial division between the expansion
of the numerator and the set of the uncut denominators.  The
coefficients of this expansion are corrected by the coefficient-level
subtractions appearing in Eq.~\eqref{eq:bubcoeffs}
and~\eqref{eq:tadcoeffs}.  The non-spurious coefficients are finally
multiplied by the corresponding Master Integrals in order to obtain
the integrated result as in Eq.~\eqref{eq:integraldecomposition}.

The coefficients of the expansions of the numerator are written on a
contiguous array by the numerator methods described in
Section~\ref{sec:input}.  The Laurent expansion is obtained via a
simplified polynomial division.  The latter is performed in-place on
the same array, keeping only the elements which are needed for the
final result.  A possible implementation for an univariate expansion,
with a numerator
\begin{equation*}
  \N = \texttt{num[0]}\, t^r + \texttt{num[1]}\, t^{r-1} + \ldots + \texttt{num[nterms-1]}\, t^{r-\texttt{nterms}+1} + \O(t^{r-\texttt{nterms}})
\end{equation*}
and denominator
\begin{equation*}
  D = \texttt{d[0]}\, t + \texttt{d[1]}\, + \texttt{d[2]}\, \frac{1}{t},
\end{equation*}
would have the form
\begin{lstlisting}
  void division(Complex num[], int nterms, Complex den[3])
  {
    for (int i=0; i<nterms; ++i) {
      num[i] /= den[0];
      if (i+1<nterms) {
        num[i+1] -= den[1]*num[i];
        if (i+2<nterms)
          num[i+2] -= den[2]*num[i];            
      }
    }
  }
\end{lstlisting}
One can check that this routine correctly replaces the first
\texttt{nterms} elements of the array \texttt{num} with the first
\texttt{nterms} leading elements of the Laurent expansion of $\N/D$.
The actual implementation in \textsc{Ninja}, having to deal with
multivariate expansions, is significantly more involved than the
\texttt{division} procedure presented here.  Nevertheless, it
qualitatively follows the same algorithm.

The coefficients obtained by the division are then corrected by the
coefficient-level subtractions and thus identified with the
corresponding coefficients of the residues, as explained in
Section~\ref{sec:laurent}.  Once the reduction is complete, the
non-spurious coefficients are multiplied by the corresponding Master
Integrals.

\subsection{Master Integrals}
\label{sec:MIs}

\textsc{Ninja} calls the routines implementing the Master Integrals
through a generic interface which, as in the case of the numerator, is
defined in the \textsc{C++} code via an abstract class (called
\texttt{IntegralLibrary}).  This allows one to use any integral
library which can be specified at run-time.  More details on the
implementation of this interface are given in
Appendix~\ref{sec:intlib}.  The current version of \textsc{Ninja}
already implements it for two integral libraries.

The first built-in interface, \texttt{ninja::AvHOneLoop}, is a
\textsc{C++} wrapper of the routines of the {\sc OneLoop} library
\cite{vanHameren:2010cp,vanHameren:2009dr}.  This wrapper caches every
computed integral allowing constant-time lookup of their values from
their arguments.  The caching of the integrals can significantly speed
up the computation, especially for complex processes.  Every instance
of the class \texttt{AvHOneLoop} has an independent cache of Master
Integrals (hence, one may safely use it in multi-threaded applications
by using one instance of the class per thread).

The second implementation of the interface, \texttt{ninja::LoopTools},
uses instead the {\sc LoopTools} library~\cite{Hahn:1998yk}, which
already has an internal cache of computed integrals.

\section{Basic usage}
\label{sec:usage}
In this section we explain how to use the library for the computation
of a generic one-loop integral.  \textsc{Ninja} can be interfaced to
any one-loop generator capable of providing the input it needs, and in
particular to packages which can reconstruct the analytic dependence
of the numerators on the loop momentum.  An interface for the one-loop
package \textsc{GoSam}~\cite{Cullen:2011ac} is already built in the
library, and has been used for highly non-trivial phenomenological
computations~\cite{vanDeurzen:2013xla,vanDeurzen:2013saa}.  An
interface with the package \textsc{FormCalc}~\cite{Hahn:1998yk} is
currently under development.  The author is open to give his
assistance in interfacing other packages as well.

In this paper we focus on the usage of \textsc{Ninja} as a standalone
library.  We will show how to generate the numerator methods needed as
input, starting from an analytic expression of the numerator, with the
help of the \textsc{Python} package \textsc{NinjaNumGen} which is
distributed with the library.  We will then explain how to perform the
reduction, and how to set the main available options.

\subsection{Installation}
\label{sec:installation}
\textsc{Ninja} can be obtained at the url
\texttt{http://ninja.hepforge.org}.  The library is distributed with
its source code using the GNU build system (also known as
\textsc{Autotools}).  It can be compiled and installed with the shell
commands
\begin{lstlisting}
   ./configure
   make
   make install
\end{lstlisting}
This will typically install the library and the header files in
sub-directories of \texttt{/usr/local}.  The \texttt{--prefix} option
can be used in order to specify a different installation path.  In
this case, you might need to update your \texttt{LD\_LIBRARY\_PATH}
(or \texttt{DYLD\_LIBRARY\_PATH} on \textsc{Mac OS}) environment
variable accordingly.  In order to use \textsc{Ninja} for the
production of phenomenological results, one must interface it with a
library of Master Integrals.  As already mentioned, interfaces to the
{\sc OneLoop} and \textsc{LoopTools} libraries are provided (see
Appendix~\ref{sec:intlib} for interfacing a different library).  They
can be enabled by passing to the \texttt{configure} script the options
\texttt{--with-avholo[=FLAGS]} and \texttt{--with-looptools[=FLAGS]}.
For instance, the following commands
\begin{lstlisting}[language=bash]
  ./configure --prefix=$\$$HOME/ninja \
      --with-avholo='-L/path/to/avh_olo/lib -lavh_olo' \
      FCINCLUDE=-I/path/to/avh_olo/include
   make install
\end{lstlisting}
will install all the files in sub-directories of \texttt{\$HOME/ninja}
and build the interface with the \textsc{OneLoop} library, which must be already installed and linkable with the flags
specified with the \texttt{--with-avholo} option.  We also specified
the \texttt{FCINCLUDE} variable with the flags which are needed to
find \textsc{Fortran-90} modules when they are not installed in a
default directory.

Given the importance of numerical precision in the calculation of
scattering amplitudes, there is also the option to compile the library
in quadruple precision (\texttt{--with-quadruple}), which uses the
\textsc{GCC libquadmath} library.  By using the \texttt{ninja::Real}
and \texttt{ninja::Complex} floating point types, the same source code
can be compiled both in double and quadruple precision, depending on
how \textsc{Ninja} has been configured.

A full list of optional arguments for the
\texttt{configure} script can be obtained with the command
\texttt{./configure --help}.  While most of them are common to every
package distributed with the GNU build system, some are instead
specific to the \textsc{Ninja} library and they are described in
Table~\ref{tab:configure}.  In most of the cases, only the options for
interfacing the integral libraries should be needed.
\begin{table}[h]
  \centering
  \begin{tabular}{| l | p{10cm} |}
    \hline
    Option & Description \\
    \hline
    \hline
    \texttt{--with-avholo[=FLAGS]} &   Include an interface with the \textsc{OneLoop} integral library~\cite{vanHameren:2010cp,vanHameren:2009dr}, specifying the corresponding flags for \emph{dynamic} linking.  If the \textsc{Fortran} module \texttt{avh\_olo} is not in a standard path, one should add its directory to the \texttt{FCINCLUDE} variable when using this option. \\
    \hline
    \texttt{--with-looptools[=FLAGS]}  & Include an interface with the \textsc{LoopTools} library~\cite{Hahn:1998yk} (needs \textsc{LoopTools} version 2.9 or higher), specifying the coresponding flags for \emph{static} linking.  If the header file \texttt{clooptools.h} is not in a standard path, one should add its directory to the \texttt{CPPFLAGS} variable when using this option. \\
    \hline
    \texttt{--with-quadruple[=FLAGS]}   &     Compile the library in quadruple precision.  This requires the \textsc{GCC libquadmath} library and one can specify the corresponding flags for the linker (\texttt{-lquadmath} by default).  With this option, the types \texttt{ninja::Real} and \texttt{ninja::Complex} will be quadruple precision floating point numbers, and including any public header file of the library will define the macro \texttt{NINJA\_QUADRUPLE} to \texttt{1} (it will not be defined otherwise).  The user should also make sure, when using this option, that the libraries of Master Integrals used by \textsc{Ninja} are compiled in quadruple precision, with floating point types compatible with the ones of \textsc{GCC}. \\
    \hline
    \texttt{--enable-higher\_rank}  &  Enable support for higher-rank numerators.  This is not needed for renormalizable theories. \\
    \hline
    \texttt{--disable-gosam}    &     Do not include \textsc{GoSam} interface. \\
    \hline
    \texttt{--disable-avholo\_cache} &  Do not include a cache of Master Integrals in the interface with the \textsc{OneLoop} library. \\
    \hline
  \end{tabular}
  \caption{Options and environment flags for the \texttt{configure} script.  Only the options which modify the default behavior of \textsc{Ninja} are listed.}
  \label{tab:configure}
\end{table}

The user can also optionally install the \textsc{Python} package
\textsc{NinjaNumGen}, which allows one to easily generate the input
needed by \textsc{Ninja}.  The package can be used both as a script
and as a \textsc{Python} module, and it could also be useful for
interfacing \textsc{Ninja} to existing one-loop packages.  In order to
install the package, move in the \texttt{utils} folder and type
\begin{lstlisting}
  python setup.py install
\end{lstlisting}
where, as usual, an installation prefix can be specified using
\texttt{--prefix}.  In this case one might need to update the
\texttt{PATH} and \texttt{PYTHONPATH} environment variables
accordingly.  The package needs
\textsc{Form-4}~\cite{Kuipers:2012rf,Kuipers:2013pba}, which should be
installed separately by the user, in order to compute the expansions
which are needed and for the generation of the corresponding source
code.

Further information about the installation procedure can be found in
the \texttt{README} file of the distribution.

\subsection{Writing the Integrand}
\label{sec:integrand}
The most important input that the \textsc{Ninja} library needs, for
the computation of a one-loop integral, is the numerator of the
corresponding integrand, cast in a suitable form.  This form has been
described in detail in Section~\ref{sec:input} but, as already
mentioned, it can also be generated with the help of the package
\textsc{NinjaNumGen}.  In this section we describe its usage as a
script, with a simple example.  A more detailed list of options,
allowing to fine-tune the output according to the user's needs, as
well as the usage of the package as a \textsc{Python} module, are
described in Appendix~\ref{sec:ninjanumgen}.

Let us define, as an example, the following 4-point one-loop
integrand, with kinematics $k_0,k_1\rightarrow k_2,k_3$
\begin{align}
  \label{eq:dummyex4}
  \I = {} & \frac{\N(q,\mu^2)}{D_0 D_1 D_2 D_3} = \frac{(q\cdot v_1)(q\cdot v_2)(q\cdot v_3)(q\cdot v_4)+\mu^4}{D_0 D_1 D_2 D_3} \nn
    D_0 ={} & \bar q^2-m_0^2\nn
    D_1 ={} & (\bar q+k_0)^2-m_1^2\nn
    D_2 ={} & (\bar q+k_0+k_1)^2-m_2^2 \nn
    D_3 ={} & (\bar q+k_3)^2-m_3^2.
\end{align}
where $v_i$ are arbitrary reference vectors.  In order to generate the
methods declared in the \texttt{ninja::Numerator} class, we first
create a file \texttt{mynum.frm} containing a
\textsc{Form}~\cite{Vermaseren:2000nd} expression for the numerator
\begin{lstlisting}
  * mynum.frm
  V v1,v2,v3,v4;
  V Q;
  S Mu2;
  
  L Diagram = (Q.v1)*(Q.v2)*(Q.v3)*(Q.v4) + Mu2^2;
\end{lstlisting}
and we run the script with the command
\begin{lstlisting}
  ninjanumgen mynum.frm --nlegs 4 --rank 4 -o mynum.cc
\end{lstlisting}
which creates the source file \texttt{mynum.cc} with the definition of
the methods, optimized for fast numerical evaluation using the recent
features of \textsc{Form-4}.  The command also creates an header file
\texttt{mynum.hh}, unless already present, which must
contain the declaration of the numerator class.  The latter will have
the following form (where for brevity we replaced the parameters of
each method with ellipses),

{\small
\begin{lstlisting}
  class Diagram : public ninja::Numerator {

  public:
    virtual ninja::Complex evaluate(...);
    virtual void muExpansion(...);
    virtual void t3Expansion(...);
    virtual void t2Expansion(...);
  
  public:
    // Add other public methods and data here

  private:
    // Add other private methods and data here
  };
\end{lstlisting}}

If the numerator expression depends on other momenta or parameters,
these should be visible inside the definitions of the methods.  In our
example, the numerator depends on the reference vectors $v_i$ which
appear in its \textsc{Form} expression.  One possibility would be
declaring these vectors as global variables, but a better alternative
could be defining them as data members of the numerator class.  In
this example we will declare them as public data members by inserting
the following code inside the class definition
\begin{lstlisting}
  public:
    ninja::ComplexMomentum v1,v2,v3,v4;
\end{lstlisting}
  which defines the vectors as complex Lorentz momenta.  This
  completes the generation of the input, which will allow
  \textsc{Ninja} to compute the integral.

\subsection{Running the reduction}
\label{sec:runtime}
In this subsection we describe the usage of \textsc{Ninja} for the
reduction of a generated integrand, such as the one in the example of
the previous subsection.  With the help of simple examples, we show
how to specify the input and how to control the run-time behavior of
the procedures of the library.  All the public header files of the
library are installed in the sub-directory \texttt{ninja} of the
\texttt{include} path in the installation directory.

\subsubsection{A simple example}
\label{sec:dummyex4}
Here we show the contents of a file \texttt{simple\_test.cc} which
illustrate the basic usage of \textsc{Ninja} for computing the
integral defined in Eq.~\eqref{eq:dummyex4}.

{\small
\begin{lstlisting}
// simple_test.cc

#include <iostream>
#include <ninja/ninja.hh>
#include <ninja/rambo.hh>
#include "mynum.hh"
using namespace ninja;

int main()
{
  // External legs of the loop
  const int N_LEGS = 4;

  // Center of mass energy
  const Real ENERGY_CM = 50;

  // Invariant s
  const Real S = ENERGY_CM*ENERGY_CM;

  // Rank of the numerator
  const int RANK = 4;

  // Declare an instance of the numerator
  Diagram num;

  // Assign numerical values to the reference vectors
  num.v1 = ComplexMomentum(1.0,1.1,1.2,1.3);
  num.v2 = ComplexMomentum(1.4,1.5,1.6,1.7);
  num.v3 = ComplexMomentum(1.8,1.9,2.0,2.1);
  num.v4 = ComplexMomentum(2.2,2.3,2.4,2.5);

  // Define external momenta
  RealMomentum k[N_LEGS];

  // Get a random phase-space point
  Rambo phase_space(S,N_LEGS);
  phase_space.getMomenta(k);

  // Define the internal momenta of the loop
  RealMomentum pi[N_LEGS];
  pi[0] = RealMomentum(0,0,0,0);
  pi[1] = k[0];
  pi[2] = k[0]+k[1];
  pi[3] = k[3];

  // Define the square of the internal masses
  Real msq[N_LEGS] = {1.,2.,3.,4.};

  // Create an amplitude object
  Amplitude<RealMasses> amp(N_LEGS,RANK,pi,msq);

  // Evaluate the integral
  amp.evaluate(num);

  // Print the result
  std::cout << amp[0] // or amp.eps0(),  finite part
            << amp[1] // or amp.epsm1(), single-pole
            << amp[2] // or amp.epsm2(), double-pole
            << std::endl;

  return 0;
}
\end{lstlisting}}

In the example above, after specifying some constants, we declare an
instance \texttt{num} of the user-defined class \texttt{Diagram},
which we constructed in Section~\ref{sec:integrand}.  Then we give
numerical values to the reference vectors $v_i$ appearing in the
numerator definition of Eq.~\eqref{eq:dummyex4} as well as in the
analytic expression given in the corresponding \textsc{Form} file.  This
defines our numerator.

Next we randomly generate a phase-space point,
by creating a \texttt{Rambo} object and calling its method
\texttt{getMomenta} which fills the array \texttt{k} of external
momenta.  The phase space generation is a translation in \textsc{C++}
of the corresponding \textsc{GoSam} implementation, which in turn is
based on the one of Ref.~\cite{Kleiss:1985gy}.  It is meant to provide an easy way to generate phase-space points in tests which use the library \textsc{Ninja}.  Every call of the
method \texttt{getMomenta} on the same \texttt{Rambo} object randomly
generates a different phase-space point.  The code above assumes the
external legs to be massless.  If the external legs were massive, with
masses
$\{\texttt{MASS\_0},\texttt{MASS\_1},\texttt{MASS\_2},\texttt{MASS\_3}\}$,
we should have generated the phase-space point by passing an array of
external masses as third argument to the constructor of the
\texttt{Rambo} object, i.e.\
\begin{lstlisting}
  Real external_masses[N_LEGS] = {MASS_0,MASS_1,
                                  MASS_2,MASS_3};
  Rambo phase_space(S,N_LEGS,external_masses);
  phase_space.getMomenta(k);
\end{lstlisting}
The method \texttt{getMomenta} can optionally take a second parameter,
namely a pointer to a \texttt{Real}, into which the weight of the
generated phase-space point will be written.  In the output, the
momenta \texttt{k[0]} and \texttt{k[1]} are taken as incoming, while
all the others are taken as outgoing.  In order to get reproducible
results, one can set the random seed with the class method
\texttt{setSeed}, which takes an integer as input.

The momenta $p_i$ are then defined according to
Eq.~\eqref{eq:dummyex4}.  Arithmetic operations between momentum types
work as one would expect.  After specifying the \emph{square} of the
internal masses, we create an \texttt{Amplitude<RealMasses>} object
\texttt{amp}, whose method \texttt{evaluate} computes the integral
with the numerator specified in its argument.  This adds the computed
integral to the total result stored internally by \texttt{amp}, which
can then be accessed either with the methods
\texttt{eps0},\texttt{epsm1},\texttt{epsm2} or with the subscript
operator ``\texttt{[]}'' as the example shows.

\subsubsection{The \texttt{Amplitude} class}
\label{sec:amplitudeclass}

The \texttt{Amplitude} template class is the main class of the
\textsc{Ninja} library and its method \texttt{evaluate} computes a
one-loop integral.  The method takes as input an object of a class
derived from \texttt{ninja::Numerator}, which provides a generic
interface to the methods defined by each Laurent expansion.  The
template parameter of the \texttt{Amplitude} class is the type of the
internal masses.  Allowed types are: \texttt{RealMasses},
\texttt{ComplexMasses} and \texttt{Massless}.  The methods needed for
the evaluation of the amplitude are instantiated for all these three
types in the compiled code of the library.

\paragraph{Instantiation}
In the example above, we showed how to instantiate an
\texttt{Amplitude} object passing to its constructor the number of
external legs, the rank of the numerator, the momenta $p_i$ and the
squared masses $m_i^2$ of the loop denominators
(Eq.~\eqref{eq:loopdenom}).  There we assumed the internal masses to
be real.  For the complex-masses case and the massless case, the
relevant part of the source would have looked like
\begin{lstlisting}
  // Complex masses
  Complex msq[N_LEGS] = {...};
  Amplitude<ComplexMasses> amp(N_LEGS,RANK,pi,msq);
  amp.evaluate(num);
\end{lstlisting}
and
\begin{lstlisting}
  // Massless (msq does not need to be specified here)
  Amplitude<Massless> amp(N_LEGS,RANK,pi);
  amp.evaluate(num);
\end{lstlisting}
respectively.  The \texttt{Amplitude} class also has a default
constructor, as well as methods which allow to set or change the
kinematics, the (squared) internal masses, the number of legs or the
rank of the numerator to be evaluated, as in the following
\begin{lstlisting}
  Amplitude<RealMasses> amp;
  amp.setN(N_LEGS);
  amp.setRank(RANK);
  amp.setKinematics(pi);
  amp.setInternalMasses(msq);
\end{lstlisting}
More in detail, the methods \texttt{setKinematics} and
\texttt{setInternalMasses} set a data member which is a pointer to the
array of momenta (\texttt{pi}) and internal masses (\texttt{msq}) to
be used respectively.  A call of one of these methods copies the
pointer given as input into the corresponding data member.  The user
should thus make sure that the pointed data will exist in memory until
the end of the execution of the \texttt{evaluate} method.

\paragraph{Renormalization scale}
  Another important setting is the renormalization scale $\mu_R^2$ to
  be used.  This is equal to 1 by default, and it only affects the
  computation of the Master Integrals.  It can be set as in the
  following example
\begin{lstlisting}
  // takes the square of the scale
  amp.setRenormalizationScale(50*50);
\end{lstlisting}

\paragraph{The S-matrix}
An optional parameter which can be set by the user is a matrix of
kinematic invariants, which we call S-matrix.  This is defined in
\textsc{Ninja} by
  \begin{equation}
    \label{eq:smat}
    s_{ij} = (p_i-p_j)^2
  \end{equation}
  where $p_i$ are the momenta appearing in Eq.~\eqref{eq:loopdenom}.
  When this is specified by the user, the computation of the Master
  Integrals might be more accurate.  This can be particularly useful
  in the presence of infrared singularities, which otherwise might not
  be detected by the integral library in use.  The S-matrix can be
  declared either by an \texttt{SMatrix} object or by a simple
  $n^2$-dimensional array, where $n$ is the number of loop
  denominators.  In the simple example given in
  Section~\ref{sec:dummyex4}, using the definitions of
  Eq.~\eqref{eq:dummyex4} in Eq.~\eqref{eq:smat}, for massless
  external momenta $k_i$ we could have specified the following
  S-matrix
  \begin{equation}
    (s_{ij}) =
    \begin{pmatrix}
      0 & 0 & 2 (k_0\cdot k_1) & 0 \\
      0 & 0 & 0 & -2(k_0\cdot k_3) \\
      2 (k_0\cdot k_1) & 0 & 0 & 0 \\
      0 & -2(k_0\cdot k_3) & 0 & 0
    \end{pmatrix}
  \end{equation}
  either with
\begin{lstlisting}
  SMatrix s_mat;
  s_mat.allocate(N_LEGS); // allocate the matrix
  s_mat.fill(0); // fill the entries with zeros
  s_mat(0,2) = s_mat(2,0) = 2*mp(k[0],k[1]);
  s_mat(1,3) = s_mat(3,1) = -2*mp(k[0],k[3]);
  amp.setSMatrix(s_mat);
\end{lstlisting}
  or with
\begin{lstlisting}
  Real s_mat[N_LEGS*N_LEGS] = {0, 0, 2*mp(k[0],k[1]), 0,
                               0, 0, 0, -2*mp(k[0],k[3]),
                               2*mp(k[0],k[1]), 0, 0, 0,
                               0, -2*mp(k[0],k[3]), 0, 0};
  amp.setSMatrix(s_mat);
\end{lstlisting}
We recommend to specify the S-Matrix whenever possible, and in
particular when infrared singularities are present.  As an alternative
to writing it explicitly, one could use the method
\begin{lstlisting}
  SMatrix &
  SMatrix::fillFromKinematics(const RealMomentum pi[],
                              Real ir_threshold = 0);
\end{lstlisting}
before each call of \texttt{evaluate}.  This will automatically
compute the matrix from the momenta $p_i$, but it will set to zero all
the matrix elements which are smaller than \texttt{ir\_threshold}.

It can be worth pointing out that an \texttt{Amplitude} object,
similarly to the case of the internal masses and the kinematics, only
stores as data member a pointer to the S-matrix (\texttt{s\_mat}) to
be used.

\paragraph{Stopping the reduction earlier} There are cases where
lower-point residues do not contribute to the final result for the
integral (see e.g.\ the example in Section~\ref{sec:sixphotons}).  If
the user knows that $k$-point residues with $k<\texttt{MIN\_CUT}$ will
not contribute to an amplitude, this information can be passed to
\textsc{Ninja} through the \texttt{setCutStop} method, as in this
example
\begin{lstlisting}
  amp.setCutStop(MIN_CUT);
\end{lstlisting}
which will tell \textsc{Ninja} to stop the reduction right after the
evaluation of the residues of $k$-ple cuts with $k=\texttt{MIN\_CUT}$.

\paragraph{Master Integrals} Each instance of an \texttt{Amplitude}
object can in principle use a different library of Master Integrals.
The library to be used can be specified with the method
\texttt{setIntegralLibrary} as in the following example
\begin{lstlisting}
  Amplitude<RealMasses> amp;
  amp.setIntegralLibrary(loop_tools);
\end{lstlisting}
If an integral library is not set explicitly for an amplitude object,
the instance will use the one which is the \emph{default} at the time
of its creation.  The default library will be \textsc{OneLoop} if
enabled during configuration, otherwise it will be \textsc{LoopTools}.
If none of the two is enabled, \textsc{Ninja} can still be used, but a
different library of Master Integrals should be specified at run-time
(more details about the implementation of the corresponding interface
are given in Appendix~\ref{sec:intlib}).  The function
\texttt{setDefaultIntegralLibrary} can be used to change the default
integral library.  Assuming both \textsc{OneLoop} and
\textsc{LoopTools} have been enabled, the user can change the default
as in the following example
\begin{lstlisting}
  #include <ninja/ninja.hh>
  #include <ninja/avholo.hh>
  #include <ninja/looptools.hh>
  using namespace ninja;

  int main()
  {
    setDefaultIntegralLibrary(loop_tools);
    // Amplitude objects defined here will use LoopTools
    setDefaultIntegralLibrary(avh_olo);
    // Amplitude objects defined here will use OneLoop
    return 0;
  }
\end{lstlisting}

\paragraph{Disabling the $\mu^2$ expansion} The user can choose to
avoid using the $\mu^2$ expansion for the boxes with
\begin{lstlisting}
  amp.useMuExpansion(false);
\end{lstlisting}
In this case the method \texttt{muExpansion} of the numerator class
does not need to be provided.  However, this would increase the number
calls of the numerator method \texttt{evaluate} and it would also
require the computation of pentagons.  Given the simplicity of the
$\mu^2$ expansion, disabling it is therefore not recommended, unless
it is done for debugging purposes.

\paragraph{Evaluation of the integrals} As already explained,
integrals are computed by calling the method \texttt{evaluate}.  This
has the following prototype
\begin{lstlisting}
  template <typename MassType>
  int Amplitude<MassType>::evaluate(Numerator & num);
\end{lstlisting}
and takes as input the numerator of the integrand, which must be an
instance of a class inherited from \texttt{ninja::Numerator}.  It
returns an integer value which depends on the results of the internal
tests which \textsc{Ninja} can optionally perform during the
computation (see Section~\ref{sec:globaloptions}).  By default no test
is performed and the return value can be safely ignored.  In general,
the return value will be equal to
\begin{description}
\item[\texttt{Amplitude<MassType>::TEST\_FAILED}] if any of the
  performed tests failed
\item[\texttt{Amplitude<MassType>::SUCCESS}] otherwise.
\end{description}
Each call of the method \texttt{evaluate} adds the computed integral
to the total result stored internally by the instance, allowing thus to easily sum different integrals with a sequence of calls of the \texttt{evaluate} method.  The total result can then be
accessed either with the methods
\texttt{eps0},\texttt{epsm1},\texttt{epsm2} or with the subscript
operator ``\texttt{[]}'' as we illustrated in the simple test
described in this section.  The result can also be quickly reset to zero,
by calling the \texttt{reset} method,
\begin{lstlisting}
  amp.reset();
\end{lstlisting}
  The finite term is given by the sum of the \emph{cut-constructible
    part} and the \emph{rational part}, but the two can also be
  accessed separately with
\begin{lstlisting}
  Complex cut_constr_part = amp.getCutConstructiblePart();
  Complex rational_part = amp.getRationalPart();
\end{lstlisting}

\subsubsection{Global settings}
\label{sec:globaloptions}
By default \textsc{Ninja} tries to compute a minimal set of
coefficients during the reduction, i.e.\ those which are needed for
the determination of the final integrated result.  These are only a
subset of the ones which are required for the full reconstruction of
the integrand decomposition of Eq.~\eqref{eq:integranddec}.  Indeed
the computation of spurious coefficients is skipped whenever possible,
i.e.\ whenever these do not enter the coefficient-level subtractions
needed for lower-point residues, as in the case of spurious
coefficients of pentagons, boxes and tadpoles.  \textsc{Ninja} also
entirely skips the computation of residues whose non-spurious
coefficients would multiply scaleless integrals.

For debugging purposes, the user can however ask \textsc{Ninja} to
perform some tests on the quality of the reconstruction of the
integrand, or print some information about the ongoing computation.
These operations might require the computation of a larger set of
coefficients.

There are two kinds of tests which \textsc{Ninja} can perform:
\emph{global tests} and \emph{local tests}.  The global $\N=\N$
test~\cite{Ossola:2006us,Ossola:2007bb} checks that the following
equality, which follows from Eq.~\eqref{eq:integranddec}, is valid
\begin{equation}
  \label{eq:neqn}
  \N(q,\mu^2) = \sum_{k=1}^5 \sum_{\{j_1,\ldots,j_k\}} \Delta_{i_1\cdots i_k}\prod_{h\neq i_1,\ldots,i_k} D_h.
\end{equation}
Another global test, which can be performed when the rank $r$ of the
numerator is equal to the number of external legs $n$, is the
so-called \emph{power-test} introduced in
Ref.~\cite{Mastrolia:2010nb}.  This consists in checking that the sum
of all the spurious tadpole coefficients is vanishing,
\begin{equation}
  \sum_{i_1=0}^{n-1} \sum_{k=0}^4 c^{(i_1)}_k = 0.
\end{equation}
Finally, \textsc{Ninja} can perform local $\N=\N$ tests on quadruple,
triple, double and single cuts.  These will check the validity of
Eq.~\eqref{eq:neqn} on values of the loop momenta corresponding to a
given $k$-ple cut, and they can be useful in order to pinpoint the
multiple cut where an error or an instability has occurred.

By default \textsc{Ninja} does not perform any internal test.  The
tests to be performed during the execution of the method
\texttt{evaluate} can be set using the function
\begin{lstlisting}
  void setTest(unsigned flag);
\end{lstlisting}
where the parameter \texttt{flag} can be any of the following:
\begin{description}
\item[\texttt{Test::NONE}] no test is performed
\item[\texttt{Test::ALL}] all tests are performed
\item[\texttt{Test::GLOBAL}] global tests are performed
\item[\texttt{Test::LOCAL\_$k$}] with $k\in\{1,2,3,4\}$, local tests on $k$-ple cuts are performed
\item[\texttt{Test::LOCAL}] all local tests are performed
\end{description}
or any combination of these.  Different flags can be combined with the
bitwise OR operator ``\texttt{|}''.  For instance, the following
command will ask \textsc{Ninja} to perform global tests and local
tests on double cuts
\begin{lstlisting}
  setTest(Test::GLOBAL | Test::LOCAL_2);
\end{lstlisting}
The $\N=\N$ tests will check whether the numerator $\N_{rec}$
reconstructed by evaluating the r.h.s.\ of Eq.~\eqref{eq:neqn} is
equal to the numerator $\N_{eva}$, obtained by a direct evaluation
through the \texttt{evaluate} method of the numerator class, up to a
given tolerance.  More explicitly, it checks if
\begin{equation}
  \left|\frac{\N_{rec}-\N_{eva}}{\N_{eva}}\right|<\delta_{tol}
\end{equation}
where the threshold $\delta_{tol}$ is $10^{-5}$ by default, but it can
specified by the user through the function
\begin{lstlisting}
  void setTestTolerance(Real test_tolerance);
\end{lstlisting}
As explained in Section~\ref{sec:amplitudeclass}, the return value of
the method \texttt{evaluate} of the \texttt{Amplitude} class can be
used in order to check whether any performed test has failed.  These
tests have been implemented for debugging purposes, and they are not
meant as an estimate of the accuracy of the total result.  Indeed,
there are cases where a numerical instability might cause a test to
fail while having negligible effects on the total amplitude.  The
accuracy of the result can be better estimated by means of the scaling
test proposed in Ref.~\cite{Badger:2010nx} or the rotation test
described in Ref.~\cite{vanDeurzen:2013saa}.

Another global option which can be set is the \emph{verbosity}, i.e.\
the amount of information printed during the evaluation of an
integral.  By default nothing is printed during a computation.  The
setting can be controlled by calling the function
\begin{lstlisting}
  void setVerbosity(unsigned flag);
\end{lstlisting}
where possible values of the parameter \texttt{flag} can be
\begin{description}
\item[\texttt{Verbose::NONE}] nothing is printed
\item[\texttt{Verbose::ALL}] everything is printed (equivalent to the combination of all the other options)
\item[\texttt{Verbose::GLOBAL\_TEST}] the result of global tests are printed, when performed
\item[\texttt{Verbose::LOCAL\_TEST\_$k$}] with $k\in\{1,2,3,4\}$, the result of local tests on $k$-ple cuts are printed when performed
\item[\texttt{Verbose::LOCAL\_TESTS}] the result of all performed local tests is printed
\item[\texttt{Verbose::TESTS}] the result of all performed tests is printed
\item[\texttt{Verbose::C$k$}] with $k\in\{1,2,3,4,5\}$, the value of the coefficients of the computed $k$-point residues is printed
\item[\texttt{Verbose::COEFFS}] the value of all the computed coefficients is printed
\item[\texttt{Verbose::RESULT}] the partial result of every call of the \texttt{evaluate} method is printed
\item[\texttt{Verbose::INTEGRALS}] the value of the Master Integrals is printed.
\end{description}
Similarly to the options controlling the performed tests, any
combination of the flags above can be specified using the bitwise OR
operator ``\texttt{|}''.  As an example, the following instruction
will ask \textsc{Ninja} to print the value of the triangle
coefficients, and the result of the current integral
\begin{lstlisting}
  setVerbosity(Verbose::C3 | Verbose::RESULT);
\end{lstlisting}
When not specified otherwise, \textsc{Ninja} will print everything to
standard output.  Any other output stream can be set, as in the
following example
\begin{lstlisting}
  #include <fstream>
  #include <ninja/ninja.hh>
  using namespace ninja;

  int main()
  {
    std::ofstream f;
    f.open("my_file.out");
    setOutputStream(f);

    // from now on everything will be printed
    // by Ninja on the file "myfile.out"

    // ...

    return 0;
  }
\end{lstlisting}

\section{Examples}
\label{sec:examples}
In this section we give a description of the examples which are
distributed with the library.  In order to compile the corresponding
executables, one can use the command
\begin{lstlisting}
  make examples
\end{lstlisting}
either in the root directory or in the \texttt{examples} directory.
These examples are meant to provide a more detailed description of the
usage of the library in several kinds of problems which involve the
computation of one-loop integrals.  More involved computations
performed with the help of \textsc{Ninja} have been presented in Ref.s
\cite{vanDeurzen:2013xla,vanDeurzen:2013saa}.  In
Ref.~\cite{vanDeurzen:2013saa} one can also find a systematic
discussion on the performance and the numerical stability of the
library in the computation of scattering amplitudes characterized by
high complexity.

Every example presented in this paper has been generated with the help
of the package \textsc{NinjaNumGen}, which is distributed with
\textsc{Ninja} and is described in Appendix~\ref{sec:ninjanumgen}.
The package can be used both as a script and as a \textsc{Python}
module.  For each example, we include in the distribution
\begin{itemize}
\item the \textsc{Form} file (with extension \texttt{.frm}) containing
the analytic expression of the numerator which is used as input
\item a \textsc{Shell} script (with extension \texttt{.sh}) with the
  command we used for the generation of the numerator class methods
\item a \textsc{Python} script (with extension \texttt{.py}) which
  achieves the same by importing the \texttt{ninjanumgen} module
\item the \textsc{C++} source files (with extension \texttt{.cc}) and
  headers files (with extension \texttt{.hh}) defining the numerator
  and its methods, as well as a test program.
\end{itemize}

\subsection{Simple Test}
\label{sec:simpletest}
The first simple example has already been extensively described in
Section~\ref{sec:usage}, in order to illustrate the basic usage of the
library.  The numerator class is defined in the header file
\texttt{mynum.hh}, the source file generated by \textsc{NinjaNumGen}
is \texttt{mynum.cc}, while the source file with the \texttt{main}
function is \texttt{simple\_test.cc}.

\subsection{Four-photon helicity amplitudes}
\label{sec:4photons}
In this example we consider a four-photon
amplitude~\cite{Gounaris:1999gh,Bernicot:2008th} and we describe the
usage of \textsc{Ninja} for the definition of polarization vectors and
other spinor objects which are needed for the evaluation of the
numerator.

The integrand of a diagram contributing to a four-photon amplitude is
given by
\begin{align}
  \I ={}& \frac{\N(q,\mu^2)}{D_0 D_1 D_2 D_3} \nn
  \N(q,\mu^2) = {}& - \textrm{Tr}\Big((\slashed{\bar l_1}+m_f)\slashed{\epsilon_1}(\slashed{\bar l_2}+m_f)\slashed{\epsilon_2}(\slashed{\bar l_3}+m_f)\slashed{\epsilon_3}(\slashed{\bar l_0}+m_f)\slashed{\epsilon_0}\Big) \nn
  D_i ={}& \bar l_i^2-m_f^2
\end{align}
where $m_f$ is the mass of the fermion propagating in the loop, and
the momenta $\bar l_i$ are defined by
\begin{align}
  \bar l_0 = \bar q, \qquad \bar l_1 = \bar q + k_0, \qquad \bar l_2 = \bar q + k_0+k_1, \qquad \bar l_3 = \bar q - k_4.
\end{align}
For simplicity we have assumed the four photons to be all incoming,
i.e.\ $k_0,k_1,k_2,k_3\rightarrow 0$.  The extra-dimensional
components $\vec{\mu}$ of the loop momentum satisfy the
(anti-)commutation relations
\begin{align}
  \{\slashed{p},\slashed{\mu}\} = 0, \qquad \{\slashed{\mu},\slashed{\mu}\} = -\mu^2,
\end{align}
for any four-dimensional momentum $p$.  This allows to work out the
extra-dimensional algebra and rewrite the numerator in terms of
four-dimensional spinor products between the polarization vectors
$\epsilon_{i}$, such as $\spa{\epsilon_i \epsilon_j}$ and
$\spb{\epsilon_i \epsilon_j}$, and scalar products involving the
four-dimensional momenta $q$, $k_i$, and momenta $e_{ij}$
defined by
\begin{align}
  \label{eq:eeij}
  e_{ij} \equiv \frac{\langle \epsilon_i \gamma^\mu \epsilon_j]}{2}.
\end{align}
The \textsc{Form} package \textsc{Spinney}~\cite{Cullen:2010jv} can
help in this kind of algebraic operations.  The final expression can
be found in the \textsc{Form} file \texttt{4photons.frm} of the
directory \texttt{examples} of the distribution.

\textsc{Ninja} includes a small library for massless spinors, which is
used internally for building the bases of momenta corresponding to
each residue.  This can also be useful for defining polarization
vectors and other spinor-related objects.  The spinor library can be
used by including the header file \texttt{ninja/spinors.hh} in the
source and linking the program with the \textsc{Ninja} library.
Polarization vectors can be defined with
\begin{lstlisting}
  // Positive helicity (right-handed)
  ComplexMomentum epsilon_r = polarizationVectorR(r,k);

  // Negative helicity (left-handed)
  ComplexMomentum epsilon_l = polarizationVectorL(r,k);
\end{lstlisting}
where \texttt{r} is an arbitrary reference momentum and \texttt{k} is
the momentum of the corresponding photon (or gluon).  These functions
assume the momenta to be incoming, while for outgoing momenta the
helicity should be reversed.  Angle-bracket and square-bracket spinor
products can be computed with the functions \texttt{spaa} and
\texttt{spbb} respectively.  If \texttt{k} is a (real or complex)
massless momentum, the corresponding spinor \texttt{spinor\_k} can be
defined as
\begin{lstlisting}
  Spinor spinor_k = Spinor(k);
\end{lstlisting}
and can be used as input parameter for the functions described above.
This turns out to be more efficient when several spinor-related
operations need to be performed on the same momentum.  Instances of
the class \texttt{Spinor} can also be used in order to define vectors
as in Eq.~\eqref{eq:eeij}, using the following function
\begin{lstlisting}
  // this returns $\(\langle p \gamma^\mu q]/2\)$
  ComplexMomentum
  momentumFromSpinors(const Spinor & p, const Spinor & q);
\end{lstlisting}

In the header file \texttt{4photons\_num.hh} we define a numerator
class \texttt{FourPhotons} containing, as private data members, the
values of all the momenta and spinor products appearing in the
integrand.  The numerator methods have been generated with
\textsc{NinjaNumGen} and written in the file
\texttt{4photons\_num.cc}.  The file \texttt{4photons\_init.cc}
contains the implementation of an \texttt{init} method which
initializes the data members using the spinor-related operations
described above, while \texttt{4photons.cc} contains a simple test.
In this test, the mass of the fermion is complex, thus it can have a
width.  The results have been compared with the ones in
Ref~\cite{Bernicot:2008th} as well as with a similar computation
performed with \textsc{Samurai} for several choices of the external
helicity states and the fermion mass.  In Figure~\ref{fig:4photonpsp}
we show a typical output for this example.
\begin{figure}[ht]
  \begin{center}
{\footnotesize
\begin{tabular}{c c c c c}
\hline 
\hline
\textsc{particle} & $E$& $p_x$ & $p_y$ & $p_z$ \\ 
\hline
    $k_0$ & 7.0000000000000000 &0.0000000000000000 &0.0000000000000000 &7.0000000000000000 \\
    $k_1$ & 7.0000000000000000 &0.0000000000000000 &                          0.0000000000000000 & -7.0000000000000000 \\
    $k_2$ & -6.9999999999999964 &-6.1126608202785198 &                          0.8284979592001092 & -3.3089226083172685 \\
    $k_3$ & -7.0000000000000027 &6.1126608202785278 &                          -0.8284979592001093 &3.3089226083172703 \\
\hline
\hline
\end{tabular}}
\vspace{0.25cm}
{\footnotesize\begin{lstlisting}[language={},showstringspaces=false]
  +----------------------------------------------------------------+
  |                                                                |
  |  Ninja - version 1.0.0                                         |
  |                                                                |
  |  Author: Tiziano Peraro                                        |
  |                                                                |
  |  Based on:                                                     |
  |                                                                |
  |      P. Mastrolia, E. Mirabella and T. Peraro,                 |
  |      "Integrand reduction of one-loop scattering amplitudes    |
  |      through Laurent series expansion,"                        |
  |      JHEP 1206 (2012) 095  [arXiv:1203.0291 [hep-ph]].         |
  |                                                                |
  |      T. Peraro,                                                |
  |      "Ninja: Automated Integrand Reduction via Laurent         |
  |      Expansion for One-Loop Amplitudes,"                       |
  |      arXiv:1403.1229 [hep-ph]                                  |
  |                                                                |
  +----------------------------------------------------------------+

Finite:      (-0.184034,-0.16765)
Abs. val.:   0.248948
Single pole: (-3.73035e-13,3.98348e-13)
Double pole: (0,0)
\end{lstlisting}}
\end{center}
  \caption{Phase space point and output for the example in \texttt{4photons.cc}.  It shows the computed finite part and poles of an all-plus four-photon helicity amplitude, using a complex fermion mass $m_f=10.0-1.0 i$.}
  \label{fig:4photonpsp}
\end{figure}

\subsection{Six-photon helicity amplitudes}
\label{sec:sixphotons}
In this example we consider six incoming
photons~\cite{Mahlon:1993fe,Nagy:2006xy,Binoth:2007ca,Ossola:2007bb,Gong:2008ww,Bernicot:2007hs,Bernicot:2008nd}.
This is a non-trivial case where the \texttt{setCutStop} method of an
\texttt{Amplitude} class can make the computation more efficient when
lower point cuts do not contribute to the total result.

A generic six-photon diagram has an integrand of the form
\begin{align}
  \I ={}& \frac{\N(q,\mu^2)}{D_0 D_1 D_2 D_3 D_4 D_5} \nn
  \N(q,\mu^2) = {}& - \textrm{Tr}\Big(\slashed{\bar l_1}\slashed{\epsilon_1}\slashed{\bar l_2}\slashed{\epsilon_2}\slashed{\bar l_3}\slashed{\epsilon_3}\slashed{\bar l_4}\slashed{\epsilon_4}\slashed{\bar l_5}\slashed{\epsilon_5}\slashed{\bar l_0}\slashed{\epsilon_0}\Big) \nn
  D_i ={}& \bar l_i^2
\end{align}
where we assumed the fermion running in the loop to be massless.  The
momenta $\bar l_i$ are defined by
\begin{alignat}{3}
  & \bar l_0 = \bar q, \qquad & & \bar l_1 = \bar q + k_0, \qquad & & \bar l_2 = \bar q + k_0+k_1, \nn
  & \bar l_3 = \bar q + k_0+k_1+k_2,  \qquad & & \bar l_2 = \bar q - k_4-k_5, \qquad & & \bar l_3 = \bar q - k_5.
\end{alignat}
One can work out the algebra, define the corresponding spinor products
and vectors, and generate the input for \textsc{Ninja} in the same way
as for the four-photons case.  One can also check that the terms
proportional to $\mu^2$ in the final expression for the integral
vanish upon integration.  Therefore, we can perform the simplification
$\bar l_i \rightarrow l_i$, or equivalently $\mu^2\rightarrow 0$, in
the numerator.  Moreover, one can exploit the knowledge that only the
cut-constructible contributions of boxes and triangles contribute to
the total result, hence we can ask \textsc{Ninja} to stop the
reduction at triple cuts with
\begin{lstlisting}
  amp.setCutStop(3);
\end{lstlisting}
and remove the rational part from the result with
\begin{lstlisting}
  amp.onlyCutConstructible();
\end{lstlisting}
which will make the computation more efficient (in the example
implemented here, the run-time is reduced by about 33\%).

In the file \texttt{6photons.cc} we call the method \texttt{evaluate}
on all the independent permutations of the external legs, generated at
run-time with the function \texttt{std::next\_permutation} of the
\textsc{C++} standard library.  The results have been compared with
the ones in Ref.s~\cite{Bernicot:2007hs,Bernicot:2008nd} as well as
with a similar computation performed with \textsc{Samurai} for several
helicity choices.

\subsection{Five-point diagram of $gg\rightarrow H t \bar t$}
With this example, we discuss a possible strategy for the generation
of the input needed by \textsc{Ninja} which can be suited for more
complex computations where an efficient evaluation of the numerator
methods at run-time can be important.

\begin{figure}[h]
  \centering
  \includegraphics[width=0.45\textwidth]{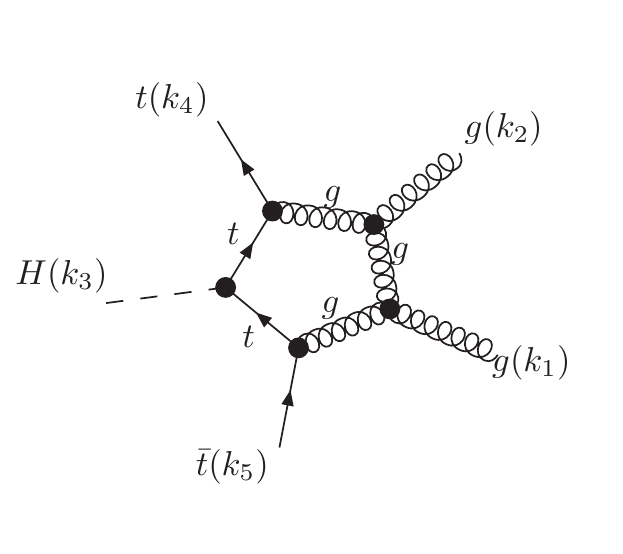}
  \caption{Diagram contributing to $gg \rightarrow H t \bar t$.  This picture has been generated using \textsc{GoSam}.}
  \label{fig:ttbarh}
\end{figure}

We consider the one-loop integral defined by the diagram depicted in
Figure~\ref{fig:ttbarh}, contributing to the 5-point helicity
amplitude $g(k_1,-), g(k_2,-)\rightarrow H(k_3), t(k_4,+), \bar t
(k_5,-)$.
\label{sec:ttbarh}
The analytic expression for the integrand of this example, which can
be worked out from the Feynman rules of the Standard Model, has been
generated with the help of the package \textsc{GoSam} and can be found
in the \textsc{Form} file \texttt{ttbarh.frm}.  This already contains
some abbreviations which are independent from the loop momentum $\bar
q$ of the diagram.  At run-time, these $\bar q$-independent
abbreviations are computed only once per phase space point, making
thus the evaluation of the numerator and its expansions more
efficient.  This analytic expression is processed by
\textsc{NinjaNumGen} which produces the numerator expansions.  We also
add to the numerator class \texttt{TTbarHDiagram} an \texttt{init}
method which uses the spinor library described in
Section~\ref{sec:4photons} in order to compute the relevant spinor
products and polarization vectors, as well as the abbreviations which
do not depend on the loop momentum.  These are stored as private data
members of the class.  For simplicity, our result neglects the
coupling constants and an overall color factor.

Even though we considered a single diagram and a specific helicity
choice, this example illustrates a general strategy for the generation
of an analytic numerator expression which is suited for the numerical
evaluations performed by integrand-reduction algorithms like the one
implemented in the library \textsc{Ninja}.  The full amplitude for
this process has been computed in
Ref.s~\cite{Beenakker:2001rj,Beenakker:2002nc,Dawson:2002tg,Dawson:2003zu,Dittmaier:2003ej},
while an additional jet has recently been added to the final state in
Ref.s~\cite{vanDeurzen:2013xla,vanDeurzen:2013saa} where
\textsc{Ninja} has been used for the reduction of the corresponding
integrands generated by \textsc{GoSam}.

\subsection{Higher-rank example}
\label{sec:hrexample}
In this example we show how \textsc{Ninja} can be used in order to
compute integrals whose rank is higher than the number of loop
denominators.  This simple test is similar to the example presented in
Ref.~\ref{sec:usage}, hence we will describe each step as in the
previous case.  We define a 5-point amplitude of rank 6, with
kinematics $k_0,k_1\rightarrow k_2,k_3,k_4,k_5$ and integrand
\begin{align}
  \I ={}& \frac{\N(q,\mu^2)}{D_0 D_1 D_2 D_3 D_4} \nn
  \N(q,\mu^2) ={}& \prod_{i=0}^2 \Big( (q\cdot v_{2i})(q\cdot v_{2i+1}) + \mu^2\, (v_{2i}\cdot v_{2i+1}) \Big) &  \nn
  D_i ={}& \bar l_i^2-m_i^2
\end{align}
in terms of the reference vectors $v_i$ ($i=0,\ldots,5$) and the
momenta $\bar l_i$ running into the loop
\begin{equation}
 \bar l_0 = \bar q, \qquad  \bar l_1 = \bar q + k_0, \qquad  \bar l_2 = \bar q + k_0+k_1,\qquad \bar l_3 = \bar q + k_3 + k_4, \qquad  \bar l_4 = \bar q + k_4.
\end{equation}

We follow the same steps outlined in Section \ref{sec:usage}.  With
\textsc{NinjaNumGen} we generate the methods for \textsc{Ninja}.
After writing the integrand in the \textsc{Form} file
\texttt{mynumhr.frm} we call the script with the command
\begin{lstlisting}
  ninjanumgen mynumhr.frm --nlegs 5 --rank 6 -o mynumhr.cc
\end{lstlisting}
which creates the file \texttt{mynumhr.cc} and a template for the
header \texttt{mynumhr.hh}.  Once again, we define the vectors $v_i$
as public members of the numerator class \texttt{Diagram}, by
inserting
\begin{lstlisting}
  public:
    ninja::ComplexMomentum v0,v1,v2,v3,v4,v5;
\end{lstlisting}
in the class definition.  A possible test program can be almost
identical to the one we showed in Section \ref{sec:dummyex4}, with
obvious changes in the definition of the rank, the number of external
legs and the reference vectors $v_i$.  This is implemented in the file
\texttt{simple\_higher\_rank\_test.cc}.  In order to run this example,
the user must compile the library with the
\texttt{--enable-higher\_rank} option, otherwise a call to the
\texttt{evaluate} method of an \texttt{Amplitude} object will cause a
run-time error.

As one can see, when \textsc{NinjaNumGen} is used for the generation
of the expansions, the higher-rank case is handled automatically
without any intervention by the user.  Besides, the internal
higher-rank routines of \textsc{Ninja} will be automatically called
whenever the rank $r$ is equal to $n+1$ (where $n$ is the number of
loop propagators), while in the public programming interface there is
no difference with respect to the normal-rank case.

\subsection{Usage in multi-threaded applications}
\label{sec:threads}
In the last examples, we wish to illustrate the possibility of using
\textsc{Ninja} in a multi-threaded application.  These examples are
implemented using \textsc{POSIX} threads, which are a standard in
Unix-like operating systems, but adapting them to different
programming interfaces for threads (such as \textsc{OpenMP}) should be
straightforward.

In order to implement a thread-safe application, one should avoid race
conditions which might occur if different threads try to write on the
same variables.  In particular, one should avoid accessing global
variables for writing from different threads.  The only global
variables used directly by \textsc{Ninja} are those controlling the
global options described in Section~\ref{sec:globaloptions}.  As
explained in that section, these options are only meant to change the
general behavior of the library for debugging purposes (e.g.\ for
checking that the provided numerator methods are correct), while in
general the default options should not be changed during a phase-space
integration, especially when performance is important.  Hence, on the
side of the \textsc{Ninja} library, there should be no issue and one
can safely call the \texttt{evaluate} method from different
\texttt{Amplitude} objects in different threads.

During a call of the \texttt{evaluate} method on an \texttt{Amplitude}
object, issues might however arise from global variables used by the
chosen library of Master Integrals or the numerator methods.  As for
the numerator methods, all the examples distributed with
\textsc{Ninja} define a thread-safe numerator class (more
specifically, one can safely call numerator methods from different
instances of the class in different threads).  This is simply done by
using data members of the class instead of global variables, making
thus different instances of the same class completely independent.

If the procedures implemented by libraries of Master Integrals are
thread-safe, one can therefore use \textsc{Ninja} in multi-threaded
applications.  As an example, one can use the class
\texttt{AvHOneLoop} which, as explained in Section~\ref{sec:MIs},
wraps routines of the \textsc{OneLoop} library and adds a cache of
computed integrals.  The cache is a non-static data member of the
class.  One can therefore create one instance of this class per thread
and assign it accordingly to the \texttt{Amplitude} objects to be
evaluated in the same thread.  As an example, with
\begin{lstlisting}
  avh_olo.init(1);
  AvHOneLoop my_lib[N_THREADS];
  Amplitude<RealMasses> amp[N_THREADS];
  for (int i=0; i<N_THREADS; ++i)
    amp[i].setIntegralLibrary(my_lib[i]);
\end{lstlisting}
we create \texttt{N\_THREADS} amplitude objects whose
\texttt{evaluate} method can be safely called in a separate thread (in
the first line, we called the \texttt{init} method on the global
instance \texttt{avh\_olo} defined in the library, in order to allow
\textsc{OneLoop} to perform its global initialization).  In this way,
different threads will also have an independent cache of Master
Integrals.  This strategy allows to build a multi-threaded application
which uses \textsc{Ninja} for the reduction of one-loop integrals.
Recent versions of \textsc{LoopTools} (namely \textsc{LoopTools-2.10}
or later) can also be used in threaded applications, since they have a
mutex regulating writing access to the internal cache of integrals.

In the following we discuss the possibility to build a multi-threaded
application with \textsc{Ninja} and any other (not necessarily
thread-safe) library of Master Integrals.  Indeed, even though
\textsc{Ninja} has obviously no control over possible issues arising
from routines of external libraries, we offer an easy way to work
around any potential problem.  In this case, there is no general way
to ensure that calling routines of the same integral library from
different threads will not cause conflicts.  However, one can avoid
these conflicts by scheduling the calls of the external procedures in
such a way that they are never evaluated at the same time from two or
more threads.  If the computation of the Master Integrals takes only a
small fraction of the total run time (which is usually the case when a
cache of integrals is present), the effects of this on the performance
will in general be reasonably small.

Within \textsc{Ninja}, implementing a scheduled access on the routines
used by a library of Master Integrals is straightforward.  As
explained more in detail in Appendix~\ref{sec:intlib}, the generic
interface used by \textsc{Ninja} in order to call Master Integral
procedures, has two methods called \texttt{init} and \texttt{exit}
which are evaluated exactly once in each call of the \texttt{evaluate}
method, immediately before the computation of the first Master
Integral and after the computation of the last Master Integral
respectively.  Therefore we can use mutexes (such as the ones present
the \textsc{POSIX} standard for threads) in order to \emph{lock} the
calls to the Master Integrals in the \texttt{init} method and
\emph{unlock} them in the \texttt{exit} method.  This makes sure that,
between the calls of the \texttt{init} and \texttt{exit} methods, no
other thread will use the Master Integral routines, hence avoiding any
possible conflict.

In order to make a library of Master Integrals thread-safe, we use the
template class \texttt{ThreadSafeIntegralLibrary}, which is included
in the distribution.  This automatically wraps an existing class
derived from \texttt{IntegralLibrary} and adds to it a mutex that
schedules the calls to the Master Integrals as explained above.  As an
example, defining a thread-safe version of a generic library
\texttt{BaseLibrary} can be simply achieved with
\begin{lstlisting}
  #include<ninja/thread_safe_integral_library.hh>
  using namespace ninja;
  ThreadSafeIntegralLibrary<BaseLibrary> my_lib;
\end{lstlisting}
which defines a new interface \texttt{my\_lib} that can be
made the default by calling
\begin{lstlisting}
  setDefaultIntegralLibrary(my_lib);
\end{lstlisting}
before any thread is created (alternatively, we could call the
\texttt{setIntegralLibrary} method on each \texttt{Amplitude} object,
either outside or inside the threads).

In the files \texttt{thread\_4photons.cc} and
\texttt{thread\_6photons.cc} we repeat the examples of the four- and
six-photons amplitudes, but this time we compute several phase-space
points in parallel on different threads.  As mentioned before, we do
not need to implement other numerator classes, since the ones
described in Sections \ref{sec:4photons} and \ref{sec:sixphotons} can
be safely used in multi-threaded applications.  In the source files,
we implement both the approaches described in this section.  The
preprocessor will select the former if the \textsc{OneLoop} interface
has been enabled and the latter otherwise.  The multi-threaded
examples can be compiled with
\begin{lstlisting}
  make thread-examples
\end{lstlisting}
if at least one between the \textsc{OneLoop} and \textsc{LoopTools}
libraries was enabled during configuration and your system supports
\textsc{POSIX} threads.

A complete discussion on the implementation of multi-threaded
applications for doing phenomenology at one-loop is beyond the
purposes of this paper.  Moreover, a detailed assessment of possible
advantages of this approach would generally depend on the generator of
the integrands and the phase space integration.  In these examples we
showed that the methods implementing the reduction via Laurent
expansion in \textsc{Ninja} can be safely used in multi-threaded
programs.

\section{Conclusions}
\label{sec:conclusions}
We presented the public \textsc{C++} library \textsc{Ninja} which
implements the Integrand Reduction via Laurent Expansion method for
the computation of one-loop amplitudes in Quantum Field Theories.  The
main procedures of the library take as input the numerator of the
integrand and some parametric expansions of the same, which can be
generated with the help of the simple \textsc{Python} package
\textsc{NinjaNumGen} included with the distribution.  The expansions
of the integrand on the multiple cuts are computed semi-numerically at
run-time, via a simplified polynomial-division algorithm.  Some of the
coefficients of the Laurent expansions are thus identified with the
ones which multiply the Master Integrals.  The algorithm is light and
proved to have good performance and numerical stability, hence it is
suited for applications to complex one-loop processes, characterized
by either several external legs or several mass scales.

We described the usage of the library, in particular the generation of
the input, the calls of the procedures for the reduction, and the
interface to libraries of Master Integrals.  This information can be
used in order to interface the library with existing one-loop
packages.  We thus expect that \textsc{Ninja} will be useful for
future computations in high-energy physics, especially for those
involving more complex processes.

\section*{Acknowledgments}
The author thanks all the other members of the \textsc{GoSam}
collaboration for the common development of a one-loop package which
could be interfaced with \textsc{Ninja}, and especially Pierpaolo
Mastrolia, Edoardo Mirabella and Giovanni Ossola for innumerable
discussions and exchanges.  The author also thanks Thomas Hahn for his
support with \textsc{LoopTools} and comments on the draft.  This work
was supported by the Alexander von Humboldt Foundation, in the
framework of the Sofja Kovalevskaja Award Project ``Advanced
Mathematical Methods for Particle Physics'', endowed by the German
Federal Ministry of Education and Research.

\begin{appendices}
\numberwithin{equation}{section}


\section{The {\sc Python} package {\sc NinjaNumGen}}
\label{sec:ninjanumgen}

The reduction procedures implemented in \textsc{Ninja} take as input a
class derived from the abstract class \texttt{ninja::Numerator}.  This
must implement the required expansions in the corresponding methods.
If the analytic expression of the numerator can be provided by the
user, the source code for the methods can be automatically generated
with the help of the simple \textsc{Python} package
\textsc{NinjaNumGen}, which is distributed with the library and can be
installed as explained in Section~\ref{sec:installation}.  The package
can be used both as a script or as a module within \textsc{Python}.

In Section \ref{sec:integrand} we already gave a simple example of its
usage as a script.  As explained there, the user must create
a file containing a \textsc{Form} expression of the numerator of the
integrand.  The package uses \textsc{Form-4} in order to generate the
expansions which are needed and produce a \textsc{C++} source file
with the definitions of the corresponding methods.  If not already
present, an header file with a sketch of the definition of the class
will also be created.  The user can complete it by adding data members
and methods which are specific of this class.  \textsc{Form} allows
one to define symbols between square brackets (e.g.\
\texttt{[symbol\_name]}), containing characters which otherwise would
not be permitted in a declaration.  \textsc{NinjaNumGen} also allows
the usage of such symbols in the expression of the numerator, and it
will remove the brackets (which would produce illegal \textsc{C++}
code) when writing the final source files.  This gives the user a
wider range of possibilities, for instance using symbols which
correspond to variable names containing underscores or data members of
structures (e.g.\ with \texttt{[structure\_instance.data\_member]}).

We first give a few more details about the usage of the package as a
script.  It is invoked with the command
\begin{lstlisting}
ninjanumgen --nlegs NLEGS ${\it optional-arguments}$ file
\end{lstlisting}
where \texttt{file} is the name of the file which contains the
numerator expression and \texttt{NLEGS} is the number of external legs
of the loop, which is equal to the number of loop denominators.  A
description of all the allowed arguments can be obtained with the
command
\begin{lstlisting}
  ninjanumgen --help
\end{lstlisting}
and the most important ones are:
\begin{description}
\item[\texttt{--rank RANK}, \texttt{-r RANK}] rank of the numerator,
  by default it will be assumed to be equal to the number of external
  legs of the loop
\item[\texttt{--diagname DIAGNAME}, \texttt{-d DIAGNAME}] name of the
  numerator expression in the \textsc{Form} file, by default it will
  be assumed to be \texttt{Diagram}
\item[\texttt{--cdiagname CDIAGNAME}] name of the numerator class in
  the generated \textsc{C++} files, by default it will be the same as
  the \textsc{Form} expression
\item[\texttt{--formexec FORMEXEC}] the \textsc{Form} executable, the
  default is \texttt{form}
\item[\texttt{--qvar QVAR}] name of the loop momentum variable $q$
  defined in Eq.~\eqref{eq:qandmu}, the default is \texttt{Q}
\item[\texttt{--mu2var MU2VAR}] name of the loop variable $\mu^2$
  defined in Eq.~\eqref{eq:qandmu}, the default is \texttt{Mu2}
\item[\texttt{--output OUTPUT}, \texttt{-o OUTPUT}] name of the output
  source file, the default is \texttt{ninjanumgen.cc}
\item[\texttt{--header HEADER}] \textsc{C++} header file containing
  the definition of the numerator class: if the file does not exists,
  one will be created.  By default it will have the same name as the
  output but with \texttt{.hh} extension.
\end{description}

As mentioned, one can also use the package as a \textsc{Python} module
(\texttt{ninjanumgen}).  This contains a class
\texttt{DiagramExpansion} which can be used for the generation of the
source code which implements the numerator methods.  The input
parameters of the constructor of this class roughly correspond to the
arguments which can be used in the script.  A detailed description can
be obtained, after installation, by invoking \textsc{Python} in
interactive mode (usually done with the command \texttt{python}) and
typing
\begin{lstlisting}[language=python]
  import ninjanumgen
  help(ninjanumgen.DiagramExpansion)
\end{lstlisting}
The method \texttt{writeSource} generates the source files.  As a
simple example, the source for the integrand we defined in
Section~\ref{sec:integrand} could have been generated within
\textsc{Python} with the commands
\begin{lstlisting}[language=python]
  # import the module
  import ninjanumgen

  # define the mandatory arguments for the constructor
  n_legs = 4
  input_file = 'mynum.frm'
  output_file = 'mynum.cc'
  
  # define an instance of the class DiagramExpansion
  mynum = ninjanumgen.DiagramExpansion(input_file,
                                       output_file,
                                       n_legs,rank=4)

  # generate the source
  mynum.writeSource()
\end{lstlisting}
We suggest to look at the \textsc{Python} files in the
\texttt{examples} directory for other basic examples.

\section{Higher-rank numerators}
\label{sec:hr}
As pointed out in Ref.~\cite{Mastrolia:2012bu}, the Laurent expansion
method can be generalized to non-renormalizable and effective theories
with higher-rank numerators.  In a renormalizable theory, with a
proper choice of gauge the rank $r$ cannot be greater than the number
$n$ of loop propagators.  \textsc{Ninja}, if configured with the
\texttt{--enable-higher\_rank} option, can also be used for the
computation of integrals with $r=n+1$.  Here we describe the
generalization of the method to the higher-rank case, underlining the
points where it differs from the renormalizable case.

In Eq.~\eqref{eq:parametricresidues}, we gave the most general
parametrization of the residues $\Delta_{i_1\cdots i_k}$ in a
renormalizable theory.  In the higher-rank case with $r=n+1$, such
parametrization is generalized as follows~\cite{Mastrolia:2012bu}
\begin{align}
  \Delta^{(r=n+1)}_{i_1 i_2 i_3 i_4 i_5} ={}& \Delta_{i_1 i_2 i_3 i_4 i_5} \nn
  \Delta^{(r=n+1)}_{i_1 i_2 i_3 i_4} ={}& \Delta_{i_1 i_2 i_3 i_4} + c_5\, \mu^4 x_{v,4}  \nn
  \Delta^{(r=n+1)}_{i_1 i_2 i_3} ={}& \Delta_{i_1 i_2 i_3} + \mu^2\, (c_{10}\, x_4^2 + c_{11}\, x_3^2) + c_{12}\, x_4^4 + c_{13}\, x_3^4 + c_{14}\, \mu^4 \nn
\Delta^{(r=n+1)}_{i_1 i_2} ={}& \Delta_{i_1 i_2} 	+ \mu^2 (  c_{10}\, x_1 + c_{11}\, x_4 + c_{12}\, x_3 )
	+ c_{13}\, x_1^3 + c_{14}\, x_4^3 + c_{15}\, x_3^3 \nn
	& + c_{16}\, x_1^2 x_4 + c_{17}\, x_1^2 x_3
	+ c_{18}\, x_1 x_4^2 + c_{19}\, x_1 x_3^2 \nn
\Delta^{(r=n+1)}_{i_1} ={}& \Delta_{i_1} + c_{5}\, x_2^2 + c_{6}\, x_1^2 + c_{7}\, x_4^2 + c_{8}\, x_3^2 + c_{10}\, x_2 x_4 + c_{11}\, x_2 x_3 \nn
    & + c_{12}\, x_1 x_4 + c_{13}\, x_1 x_3
	+ c_{14}\, \mu^2 + c_{15}\, x_3 x_4.
\label{eq:hrparametricresidues}
\end{align}
The generalized integral decomposition is thus
\begin{align}
  \M^{(r=n+1)} = {}& \M^{(r=n)}+ 
\sum_{\{i_1, i_2, i_3\}}\, 
          c_{14}^{ (i_1 i_2 i_3)} I_{i_1 i_2 i_3}[\mu^4]
\pagebreak[1] \nn
     & +
\sum_{\{i_1, i_2\}}\bigg\{
          c_{10}^{ (i_1 i_2)} I_{i_1 i_2}[\mu^2\, (q+p_{i_1})\cdot e_2) ]
        + c_{13}^{ (i_1 i_2)} I_{i_1 i_2}[((q+p_{i_1})\cdot e_2)^3 ] \bigg\} \pagebreak[1] \nn 
& + 
\sum_{i_1}
     \bigg\{ c_{14}\, I_{i_1}[\mu^2] + c_{15}^{ (i_1)}\, I_{i_1}[((q+p_{i_1})\cdot e_3)((q+p_{i_1})\cdot e_4)] \bigg\}  \label{eq:hrintegraldecomposition}
\end{align}
This higher-rank decomposition has been used for the computation of
NLO corrections to Higgs-boson production in association with two
\cite{vanDeurzen:2013rv} and three
\cite{Cullen:2013saa,vanDeurzen:2013saa} jets.  Other libraries which
implement the reduction of higher-rank integrals are
\textsc{Xsamurai}~\cite{vanDeurzen:2013pja}, which extends the more
traditional integrand reduction algorithm of \textsc{Samurai}, and
\textsc{Golem95}~\cite{Binoth:2008uq,Cullen:2011kv,Guillet:2013msa}.

\subsection{Reduction of higher-rank integrands}
While the extension of the Laurent expansion method for the
computation of higher-rank 3-point and 2-point residues is
straightforward, for 4-point and 1-point residues some further
observations are in order.  Here we propose a generalization of the
Laurent expansion method which allows to efficiently compute the
non-spurious coefficients of 4- and 1-point residues without spoiling
the nice features of the algorithm, such as the simplified
subtractions of higher-point contributions and the diagonal systems of
equations.  This generalization is not present elsewhere in the
literature and has been implemented in the \textsc{Ninja} library.

\paragraph{4-point residues}
The coefficient $c_0$ can be computed exactly as in the renormalizable
case.  For the coefficient $c_{4}$, one needs instead to keep also the
next-to-leading term in the $\mu^2$ expansion described before, so
that the $d$-dimensional solutions of a quadruple cut, given in
Eq.~\eqref{eq:qmuexp}, in the asymptotic limit become
\begin{equation}
  \label{eq:qmuexphr}
  q_{\pm}^\nu = -p_{i_1}^\nu + a^\nu \pm \sqrt{\mu^2+\beta}\, v_\perp^\nu\; \overset{\mu^2\rightarrow\infty}{=}\; -p_{i_1}^\nu +  a^\nu \pm \sqrt{\mu^2} \, v_\perp^\nu + \O\left(\frac{1}{\sqrt{\mu^2}}\right),
\end{equation}
where it is worth noticing that $a^\nu$ can be obtained as the average
of the two solutions of the corresponding four-dimensional quadruple
cut.  In this limit, the expansion of the integrand reads
\begin{equation}
  \left. \frac{\N(q,\mu^2)}{\prod_{j\neq i_1,i_2,i_3,i_4}D_j}
  \right|_{q=\sqrt{\mu^2} v_\perp+a+
    \O(\mu^{-1})} = c_5\, v_{\perp}^2\, \mu^5 + c_4 \mu^4 + \O(\mu^3),
\end{equation}
hence the leading term is now the spurious coefficient
$c_5$, but $c_4$ can still be obtained as the next-to-leading term.  This can
be implemented semi-numerically, by keeping the two leading terms of
the expansion of the numerator and performing a polynomial division
with respect to the two leading terms in the expansion of the uncut
denominators which have the form
\begin{equation}
  \left. D_{h\neq i_1, i_2, i_3, i_4}\right|_{q=q_+} = d_{h,0}\, \sqrt{\mu^2}+ d_{h,1}+\O\left(\frac{1}{\sqrt{\mu^2}}\right).
\end{equation}
Given the very limited number of terms involved, the division can be
implemented very efficiently in a small number of operations.  More in
detail, if \texttt{num} and \texttt{den} are arrays of length two
containing the leading and next-to-leading terms in the expansion of
the numerator and a denominator respectively, we can perform the
division in place with the commands
\begin{lstlisting}
   num[0] /= den[0];
   num[1] -= den[1]*num[0];
   num[1] /= den[0];
\end{lstlisting}
which will have the effect of replacing the entries of \texttt{num}
with the ones of the expansion of $\N/D$.
We also observe that the computation and the subtraction of pentagons is
not needed in the higher-rank case either.

\paragraph{1-point residues}
On higher-rank 1-point residues $\Delta_{i_1}$ we consider
$d$-dimensional solutions of the corresponding single cut of the form
\begin{equation}
  q_+^\nu = -p_{i_1}+t\, e_1^\nu+\frac{m_{i_1}^2+\mu^2}{2\, t\, (e_1\cdot e_2)}\, e_2^\nu,
 \qquad   q_-^\nu =-p_{i_1}+ t\, e_3^\nu+\frac{m_{i_1}^2+\mu^2}{2\, t\, (e_3
   \cdot e_4)}\, e_4^\nu,
\end{equation}
in terms of the free variables $t$ and $\mu^2$.  By taking the
$t\rightarrow\infty$ limit of the integrand and the subtraction terms
evaluated on these solutions, we obtain an asymptotic polynomial
expansion of the form
\begin{align}
  \frac{\N(q_\pm,\mu^2)}{\prod_{j\neq i_1}D_j} & = n^{\pm}_0 + n^{\pm}_1\, t
  + n^{\pm}_{2} \, t^2 + n^{\pm}_{3}\, \mu^2 + \O(1/t) \\
  \frac{\Delta_{i_1 j}(q_\pm,\mu^2)}{D_j} & = c_{s_2,0}^{\pm(j)} + c_{s_2,1}^{\pm(j)}\, t + c_{s_2,2}^{\pm(j)}\, t^2 + c_{s_2,3}^{\pm(j)}\, \mu^2 + \O(1/t) \\
  \frac{\Delta_{i_1 j k}(q_\pm,\mu^2)}{D_j D_k} & = c_{s_3,0}^{\pm(jk)} + c_{s_3,1}^{\pm(jk)}\, t + c_{s_3,2}^{\pm(jk)}\, t^2 + c_{s_3,3}^{\pm(jk)}\, \mu^2 + \O(1/t).
\end{align}
One can check that the non-spurious coefficients of the tadpole are
given in terms of the ones of the expansions above by
\begin{align}
  c_0 & = n^+_0 - \sum_j c_{s_2,0}^{+(j)}  - \sum_{jk}
  c_{s_3,0}^{+(jk)}, \nn
  c_{14} & = n^+_3 - \sum_j c_{s_2,3}^{+(j)} - \sum_{jk}
c_{s_3,3}^{+(jk)}, \nn
  c_{15} & = \frac{2}{(e_3\cdot e_4)}\Big(n^-_3 - \sum_j c_{s_2,3}^{-(j)} - \sum_{jk}
c_{s_3,3}^{-(jk)} - c_{14}\Big).
\end{align}

\subsection{The input}
In the higher-rank case, the \texttt{muExpansion} method of the
numerator needs to compute both the leading and the next-to-leading
term of the expansion in $\mu^2$.  The package \textsc{NinjaNumGen},
takes care of this automatically when the specified rank is higher
than the number of external legs of the loop.  The information in the
next paragraph is only needed for a custom implementation of the
method without \textsc{NinjaNumGen}.

The \texttt{muExpansion} method in the higher-rank case should compute
the two leading terms of the expansion in $t$ of the numerator,
defined by
  \begin{equation}
    q^\nu \rightarrow t\, v_0^\nu\, t+v_1^\nu ,\qquad \mu^2 \rightarrow t^2\, v_0^2
  \end{equation}
  with $v_i\equiv v\texttt{\texttt{[i]}}$, where $v$ is the array of
  momenta passed as input parameter.  The leading and next-to-leading
  terms of the expansion should be written in the first two entry of
  the array pointed by the parameter \texttt{c}, i.e.\
\begin{lstlisting}
    c[0] = $\(\N\)$[$\(t^{r}\)$];
    c[1] = $\(\N\)$[$\(t^{r-1}\)$];
\end{lstlisting}
All the other methods should have instead the same definition
described in Section~\ref{sec:input}.

\subsection{Higher-rank Master Integrals}
\label{sec:hrmis}
As one can see from Eq.~\eqref{eq:hrintegraldecomposition}, in the
higher-rank case five new types of integral appear in the final
decomposition.  They are a 2-point integral of rank 3, a 1-point
integral of rank 2, and three more integrals containing $\mu^2$ at the
numerator which contribute to the rational part of the amplitude.

\textsc{Ninja} contains an implementation of all these higher-rank
integrals in terms of lower-rank ones.  This means that, should the
user choose to interface a custom integral library (see
Appendix~\ref{sec:intlib}), these higher-rank integrals would not be
needed, although specifying an alternative implementation would still
be possible.

All the integrals of Eq.~\eqref{eq:hrintegraldecomposition} which
contribute to the rational part of the amplitude have already been
computed in Ref.~\cite{Mastrolia:2012bu}.  With our choice of the
normalization factor given in Eq.~\eqref{eq:hmurd}, they read
\begin{align}
  I_{i_1 i_2 i_3}[\mu^4] ={}& \frac{1}{6}\left( \frac{s_{i_2 i_1} + s_{i_3 i_2} + s_{i_1 i_3}}{4} - m_{i_1}^2 - m_{i_2}^2 - m_{i_3}^2 \right) +\O(\epsilon)\\
  I_{i_1 i_2}[\mu^2\, ((q+p_{i_1}) \cdot v)] ={}& \frac{((p_{i_2}-p_{i_1}) \cdot v)}{12}\Big( s_{i_2 i_1} - 2\, m_{i_1}^2 - 4\, m_{i_2}^2 \Big)  +\O(\epsilon) \\
  I_{i_1}[\mu^2] ={}& \frac{m_{i_1}^4}{2}   +\O(\epsilon)
\end{align}
where $s_{ij}$ were defined in Eq.~\eqref{eq:smat} and $v$ is an
arbitrary vector.  The tadpole of rank 2 appearing in
Eq.~\eqref{eq:hrintegraldecomposition} can also be written as a
function of the scalar tadpole integral $I_{i_1}$ as follows
\begin{equation}
  I_{i_1}[((q+p_{i_1})\cdot e_3)\, ((q+p_{i_1})\cdot e_4)] ={} m_{i_1}^2\, \frac{(e_3\cdot e_4)}{4}\left( I_{i_1} + \frac{m_{i_1}^2}{2} \right)   + \O(\epsilon).
\end{equation}
Since the vector $e_{2}$ in the bubble integral of rank 3 appearing in
Eq.~\eqref{eq:hrintegraldecomposition} is massless, the corresponding
integral is simply proportional to the form factor $B_{111}$,
\begin{equation}
  I_{i_1 i_2}[((q+p_{i_1})\cdot e_2)^3 ] = ((p_{i_2}-p_{i_1})\cdot e_2)^3\, B_{111}(s_{i_2 i_1},m_{i_1}^2, m_{i_2}^2).
\end{equation}
The form factor can be computed using the formulas of
Ref.~\cite{Stuart:1987tt}, as a function of form factors of scalar
integrals $B_0$.  In the special case with $s_{i_2 i_1} = 0$ we use
Eq.~(A.6.2) and (A.6.3) of that reference.  For the general case
$s_{i_2 i_1} \neq 0$ we implement instead the following formula
\begin{align}
  B_{111}(s_{i_2 i_1},m_{i_1}^2,m_{i_2}^2) ={}\frac{1}{4\, s_{i_2 i_1}^3} \bigg\{& s_{i_2 i_1}\, \Big(m_{i_1}^2 
\, I_{i_1}+ I_{i_1}[\mu^2]-m_{i_2}^2 
\, I_{i_2}-I_{i_2}[\mu^2] \nn & - 4 
\, I_{i_1 i_2}[\mu^2\, ((q+p_{i_1})\cdot (p_{i_2}-p_{i_1}))] \nn & - 4 \, m_{i_1}^2
\, I_{i_1 i_2}[(q+p_{i_1})\cdot (p_{i_2}-p_{i_1})]\Big) \nn
& + 4\, (m_{i_2}^2-m_{i_1}^2-s_{i_2 i_1}) \, I_{i_1 i_2}[((q+p_{i_1})\cdot (p_{i_2}-p_{i_1}))^2] \bigg\}.
\end{align}

\section{Interfaces to Integral Libraries}
\label{sec:intlib}
\textsc{Ninja} already implements interfaces for the \textsc{OneLoop}
and the \textsc{LoopTools} integral libraries.  These libraries have
been used in a large number of computations and provide very reliable
results, hence they should suffice for most purposes.  However,
\textsc{Ninja} has been designed considering the possibility of using
any other library of Master Integrals.

The Master Integrals are computed by calling virtual methods of the
abstract class \texttt{ninja::IntegralLibrary}, which is defined in
the header file \texttt{ninja/integral\_library.hh}.  Therefore, any
library of Master Integrals can be interfaced by implementing a class
derived from \texttt{IntegralLibrary}.  Each method of the library
corresponds to a different Master Integral appearing in
Eq.~\eqref{eq:integraldecomposition}, which should be implemented for
both real and complex internal masses (and optionally for the massless
case).  An implementation of higher-rank integrals can also be
provided but it is not needed, since \textsc{Ninja} has a default
implementation of them in terms of lower rank integrals.  There are
two further methods, namely \texttt{init} and \texttt{exit}.  The
former is called inside the method \texttt{Amplitude::evaluate} just
before the computation of the first needed Master Integral, while the
latter is called after the last Master Integral has been computed.
The method \texttt{init}
\begin{lstlisting}
  	virtual void init(Real muRsq) = 0;
\end{lstlisting}
takes as input the square of the renormalization scale to be used in
the subsequent calls of the methods implementing the Master Integrals.
It can also be used in order to perform any other initialization the
library might need before computing the integrals.  The \texttt{exit}
method instead, does not need to be implemented and by default it will
not perform any action.  In Section~\ref{sec:threads} we gave an
example of a case where a non-trivial implementation of the
\texttt{exit} method could be useful.

The other methods should compute the finite part and the poles of the
corresponding Master Integrals.  As an example, the methods for the
box integrals have the following declarations
\begin{lstlisting}
  // - real masses
  virtual void
  getBoxIntegralRM(Complex rslt[3],
                   Real s21, Real s32, Real s43,
                   Real s14, Real s31, Real s42,
                   Real m1sq, Real m2sq,
                   Real m3sq, Real m4sq) = 0;
 
  // - complex masses
  virtual void
  getBoxIntegralCM(Complex rslt[3],
                   Real s21, Real s32, Real s43,
                   Real s14, Real s31, Real s42,
                   const Complex & m1sq,
                   const Complex & m2sq,
                   const Complex & m3sq,
                   const Complex & m4sq) = 0;
\end{lstlisting}
and they must write the $\O(\epsilon^{-i})$ term of the
result in the $i$-th entry of the array \texttt{rslt}, for
$i\in\{0,1,2\}$.  The arguments are the invariants $s_{ij}$ and the
squared masses $m_i^2$.  Similar methods need to be provided for
3-point, 2-point and 1-point Master Integrals, as described in detail
in the comments inside the header file
\texttt{ninja/integral\_library.hh}.

\subsection{Built-in interfaces}
\label{sec:misbuiltin}
Examples of implementation of this interface for the libraries
\textsc{OneLoop} and \textsc{LoopTools} can be found in the source
code.  More in detail, we define the instances
\texttt{ninja::avh\_olo} and \texttt{ninja::loop\_tools} of the
classes \texttt{ninja::AvHOneLoop} and \texttt{ninja::LoopTools}
respectively, which implement the methods described above as wrappers
of the corresponding routines in each integral library.

The \textsc{OneLoop} interface also implements a cache of Master
Integrals on top of these routines.  The cache is implemented
similarly to a hash table, which allows constant-time look-up of each
computed integral from its arguments.  Hence, the methods of the
\texttt{AvHOneLoop} class will call the routines of the
\textsc{OneLoop} library only if a Master Integral is not found in the
cache.  The cache can be cleared with the class method
\texttt{AvHOneLoop::clearIntegralCache}.  During a phase-space
integration, we suggest calling this method once per phase space
point, especially for more complex processes.  This method does not
completely free the allocated memory, but keeps the buckets of the
hash table available in order to store the integrals more efficiently
in subsequent calls of the respective methods.  If the user wishes to
completely free the allocated memory, the method
\texttt{AvHOneLoop::freeIntegralCache} can be used, although in
general \texttt{clearIntegralCache} should be preferred.  As already
mentioned, every instance of \texttt{AvHOneLoop} has a cache of Master
Integrals as data member.  This can be useful for building
multi-threaded applications, as discussed in the examples of
Section~\ref{sec:threads}.

Since \textsc{LoopTools} already has an internal cache of Master
Integrals, the implementation of its interface is much simpler and
only consists in wrapper of its routines.  We implemented a
\texttt{clearIntegralCache} method in the \texttt{LoopTools} class as
well, which in this case simply calls the routine which clears the
cache of integrals in \textsc{LoopTools}.

\end{appendices}


\begin{thebibliography}{99}
\bibitem{Cachazo:2004kj}
  F.~Cachazo, P.~Svrcek and E.~Witten,
  JHEP {\bf 0409} (2004) 006
  [hep-th/0403047].



\bibitem{Britto:2004ap}
  R.~Britto, F.~Cachazo and B.~Feng,
  Nucl.\ Phys.\ B {\bf 715} (2005) 499
  [hep-th/0412308].



\bibitem{Britto:2005fq}
  R.~Britto, F.~Cachazo, B.~Feng and E.~Witten,
  Phys.\ Rev.\ Lett.\  {\bf 94} (2005) 181602
  [hep-th/0501052].



\bibitem{Bern:1994zx}
  Z.~Bern, L.~J.~Dixon, D.~C.~Dunbar and D.~A.~Kosower,
  Nucl.\ Phys.\ B {\bf 425} (1994) 217
  [hep-ph/9403226].



\bibitem{Britto:2004nc}
  R.~Britto, F.~Cachazo and B.~Feng,
  Nucl.\ Phys.\ B {\bf 725} (2005) 275
  [hep-th/0412103].



\bibitem{Ossola:2006us}
  G.~Ossola, C.~G.~Papadopoulos and R.~Pittau,
  Nucl.\ Phys.\ B {\bf 763} (2007) 147
  [hep-ph/0609007].



\bibitem{Ellis:2007br}
  R.~K.~Ellis, W.~T.~Giele and Z.~Kunszt,
  JHEP {\bf 0803} (2008) 003
  [arXiv:0708.2398 [hep-ph]].



\bibitem{Mastrolia:2011pr}
  P.~Mastrolia and G.~Ossola,
  JHEP {\bf 1111} (2011) 014
  [arXiv:1107.6041 [hep-ph]].



\bibitem{Badger:2012dp}
  S.~Badger, H.~Frellesvig and Y.~Zhang,
  JHEP {\bf 1204} (2012) 055
  [arXiv:1202.2019 [hep-ph]].



\bibitem{Zhang:2012ce}
  Y.~Zhang,
  JHEP {\bf 1209} (2012) 042
  [arXiv:1205.5707 [hep-ph]].



\bibitem{Mastrolia:2012an}
  P.~Mastrolia, E.~Mirabella, G.~Ossola and T.~Peraro,
  Phys.\ Lett.\ B {\bf 718} (2012) 173
  [arXiv:1205.7087 [hep-ph]].



\bibitem{Mastrolia:2013kca}
  P.~Mastrolia, E.~Mirabella, G.~Ossola and T.~Peraro,
  Phys.\ Lett.\ B {\bf 727} (2013) 532
  [arXiv:1307.5832 [hep-ph]].



\bibitem{Ossola:2007ax}
  G.~Ossola, C.~G.~Papadopoulos and R.~Pittau,
  JHEP {\bf 0803} (2008) 042
  [arXiv:0711.3596 [hep-ph]].



\bibitem{Mastrolia:2010nb}
  P.~Mastrolia, G.~Ossola, T.~Reiter and F.~Tramontano,
  JHEP {\bf 1008} (2010) 080
  [arXiv:1006.0710 [hep-ph]].



\bibitem{Hahn:1998yk}
  T.~Hahn and M.~Perez-Victoria,
  Comput.\ Phys.\ Commun.\  {\bf 118} (1999) 153
  [hep-ph/9807565].



\bibitem{vanHameren:2009dr}
  A.~van Hameren, C.~G.~Papadopoulos and R.~Pittau,
  JHEP {\bf 0909} (2009) 106
  [arXiv:0903.4665 [hep-ph]].



\bibitem{Bevilacqua:2011xh}
  G.~Bevilacqua, M.~Czakon, M.~V.~Garzelli, A.~van Hameren, A.~Kardos, C.~G.~Papadopoulos, R.~Pittau and M.~Worek,
  Comput.\ Phys.\ Commun.\  {\bf 184} (2013) 986
  [arXiv:1110.1499 [hep-ph]].



\bibitem{Berger:2008sj}
  C.~F.~Berger, Z.~Bern, L.~J.~Dixon, F.~Febres Cordero, D.~Forde, H.~Ita, D.~A.~Kosower and D.~Maitre,
  Phys.\ Rev.\ D {\bf 78} (2008) 036003
  [arXiv:0803.4180 [hep-ph]].



\bibitem{Hirschi:2011pa}
  V.~Hirschi, R.~Frederix, S.~Frixione, M.~V.~Garzelli, F.~Maltoni and R.~Pittau,
  JHEP {\bf 1105} (2011) 044
  [arXiv:1103.0621 [hep-ph]].



\bibitem{Cullen:2011ac}
  G.~Cullen, N.~Greiner, G.~Heinrich, G.~Luisoni, P.~Mastrolia, G.~Ossola, T.~Reiter and F.~Tramontano,
  Eur.\ Phys.\ J.\ C {\bf 72} (2012) 1889
  [arXiv:1111.2034 [hep-ph]].



\bibitem{Cascioli:2011va}
  F.~Cascioli, P.~Maierhofer and S.~Pozzorini,
  Phys.\ Rev.\ Lett.\  {\bf 108} (2012) 111601
  [arXiv:1111.5206 [hep-ph]].



\bibitem{Badger:2010nx}
  S.~Badger, B.~Biedermann and P.~Uwer,
  Comput.\ Phys.\ Commun.\  {\bf 182} (2011) 1674
  [arXiv:1011.2900 [hep-ph]].



\bibitem{Badger:2012pg}
  S.~Badger, B.~Biedermann, P.~Uwer and V.~Yundin,
  Comput.\ Phys.\ Commun.\  {\bf 184} (2013) 1981
  [arXiv:1209.0100 [hep-ph]].



\bibitem{Mastrolia:2012bu}
  P.~Mastrolia, E.~Mirabella and T.~Peraro,
  JHEP {\bf 1206} (2012) 095
   [Erratum-ibid.\  {\bf 1211} (2012) 128]
  [arXiv:1203.0291 [hep-ph]].



\bibitem{vanHameren:2010cp}
  A.~van Hameren,
  Comput.\ Phys.\ Commun.\  {\bf 182} (2011) 2427
  [arXiv:1007.4716 [hep-ph]].



\bibitem{vanDeurzen:2013xla}
  H.~van Deurzen, G.~Luisoni, P.~Mastrolia, E.~Mirabella, G.~Ossola and T.~Peraro,
  Phys.\ Rev.\ Lett.\  {\bf 111} (2013) 171801
  [arXiv:1307.8437 [hep-ph]].



\bibitem{vanDeurzen:2013saa}
  H.~van Deurzen, G.~Luisoni, P.~Mastrolia, E.~Mirabella, G.~Ossola and T.~Peraro,
  arXiv:1312.6678 [hep-ph].



\bibitem{Vermaseren:2000nd}
  J.~A.~M.~Vermaseren,
  math-ph/0010025.



\bibitem{Kuipers:2012rf}
  J.~Kuipers, T.~Ueda, J.~A.~M.~Vermaseren and J.~Vollinga,
  Comput.\ Phys.\ Commun.\  {\bf 184} (2013) 1453
  [arXiv:1203.6543 [cs.SC]].



\bibitem{Kuipers:2013pba}
  J.~Kuipers, T.~Ueda and J.~A.~M.~Vermaseren,
  arXiv:1310.7007 [cs.SC].



\bibitem{Ellis:2007qk}
  R.~K.~Ellis and G.~Zanderighi,
  JHEP {\bf 0802} (2008) 002
  [arXiv:0712.1851 [hep-ph]].



\bibitem{Forde:2007mi}
  D.~Forde,
  Phys.\ Rev.\ D {\bf 75} (2007) 125019
  [arXiv:0704.1835 [hep-ph]].



\bibitem{delAguila:2004nf}
  F.~del Aguila and R.~Pittau,
  JHEP {\bf 0407} (2004) 017
  [hep-ph/0404120].



\bibitem{Mastrolia:2008jb}
  P.~Mastrolia, G.~Ossola, C.~G.~Papadopoulos and R.~Pittau,
  JHEP {\bf 0806} (2008) 030
  [arXiv:0803.3964 [hep-ph]].



\bibitem{Badger:2008cm}
  S.~D.~Badger,
  JHEP {\bf 0901} (2009) 049
  [arXiv:0806.4600 [hep-ph]].



\bibitem{Kleiss:1985gy}
  R.~Kleiss, W.~J.~Stirling and S.~D.~Ellis,
  Comput.\ Phys.\ Commun.\  {\bf 40} (1986) 359.



\bibitem{Ossola:2007bb}
  G.~Ossola, C.~G.~Papadopoulos and R.~Pittau,
  JHEP {\bf 0707} (2007) 085
  [arXiv:0704.1271 [hep-ph]].



\bibitem{Gounaris:1999gh}
  G.~J.~Gounaris, P.~I.~Porfyriadis and F.~M.~Renard,
  Eur.\ Phys.\ J.\ C {\bf 9} (1999) 673
  [hep-ph/9902230].



\bibitem{Bernicot:2008th}
  C.~Bernicot,
  arXiv:0804.0749 [hep-ph].



\bibitem{Cullen:2010jv}
  G.~Cullen, M.~Koch-Janusz and T.~Reiter,
  Comput.\ Phys.\ Commun.\  {\bf 182} (2011) 2368
  [arXiv:1008.0803 [hep-ph]].



\bibitem{Mahlon:1993fe}
  G.~Mahlon,
  Phys.\ Rev.\ D {\bf 49} (1994) 2197
  [hep-ph/9311213].



\bibitem{Nagy:2006xy}
  Z.~Nagy and D.~E.~Soper,
  Phys.\ Rev.\ D {\bf 74} (2006) 093006
  [hep-ph/0610028].



\bibitem{Binoth:2007ca}
  T.~Binoth, G.~Heinrich, T.~Gehrmann and P.~Mastrolia,
  Phys.\ Lett.\ B {\bf 649} (2007) 422
  [hep-ph/0703311].



\bibitem{Gong:2008ww}
  W.~Gong, Z.~Nagy and D.~E.~Soper,
  Phys.\ Rev.\ D {\bf 79} (2009) 033005
  [arXiv:0812.3686 [hep-ph]].



\bibitem{Bernicot:2007hs}
  C.~Bernicot and J.~-P.~.Guillet,
  JHEP {\bf 0801} (2008) 059
  [arXiv:0711.4713 [hep-ph]].



\bibitem{Bernicot:2008nd}
  C.~Bernicot,
  arXiv:0804.1315 [hep-ph].



\bibitem{Beenakker:2001rj}
  W.~Beenakker, S.~Dittmaier, M.~Kramer, B.~Plumper, M.~Spira and P.~M.~Zerwas,
  Phys.\ Rev.\ Lett.\  {\bf 87} (2001) 201805
  [hep-ph/0107081].



\bibitem{Beenakker:2002nc}
  W.~Beenakker, S.~Dittmaier, M.~Kramer, B.~Plumper, M.~Spira and P.~M.~Zerwas,
  Nucl.\ Phys.\ B {\bf 653} (2003) 151
  [hep-ph/0211352].



\bibitem{Dawson:2002tg}
  S.~Dawson, L.~H.~Orr, L.~Reina and D.~Wackeroth,
  Phys.\ Rev.\ D {\bf 67} (2003) 071503
  [hep-ph/0211438].



\bibitem{Dawson:2003zu}
  S.~Dawson, C.~Jackson, L.~H.~Orr, L.~Reina and D.~Wackeroth,
  Phys.\ Rev.\ D {\bf 68} (2003) 034022
  [hep-ph/0305087].



\bibitem{Dittmaier:2003ej}
  S.~Dittmaier, M.~Kramer, 1 and M.~Spira,
  Phys.\ Rev.\ D {\bf 70} (2004) 074010
  [hep-ph/0309204].



\bibitem{vanDeurzen:2013rv}
  H.~van Deurzen, N.~Greiner, G.~Luisoni, P.~Mastrolia, E.~Mirabella, G.~Ossola, T.~Peraro and J.~F.~von Soden-Fraunhofen {\it et al.},
  Phys.\ Lett.\ B {\bf 721} (2013) 74
  [arXiv:1301.0493 [hep-ph]].



\bibitem{Cullen:2013saa}
  G.~Cullen, H.~van Deurzen, N.~Greiner, G.~Luisoni, P.~Mastrolia, E.~Mirabella, G.~Ossola and T.~Peraro {\it et al.},
  Phys.\ Rev.\ Lett.\  {\bf 111} (2013) 131801
  [arXiv:1307.4737 [hep-ph]].



\bibitem{vanDeurzen:2013pja}
  H.~van Deurzen,
  Acta Phys.\ Polon.\ B {\bf 44} (2013) 11,  2223.



\bibitem{Binoth:2008uq}
  T.~Binoth, J.~-P.~.Guillet, G.~Heinrich, E.~Pilon and T.~Reiter,
  Comput.\ Phys.\ Commun.\  {\bf 180} (2009) 2317
  [arXiv:0810.0992 [hep-ph]].



\bibitem{Cullen:2011kv}
  G.~Cullen, J.~P.~.Guillet, G.~Heinrich, T.~Kleinschmidt, E.~Pilon, T.~Reiter and M.~Rodgers,
  Comput.\ Phys.\ Commun.\  {\bf 182} (2011) 2276
  [arXiv:1101.5595 [hep-ph]].



\bibitem{Guillet:2013msa}
  J.~P.~.Guillet, G.~Heinrich and J.~F.~von Soden-Fraunhofen,
  arXiv:1312.3887 [hep-ph].



\bibitem{Stuart:1987tt}
  R.~G.~Stuart,
  Comput.\ Phys.\ Commun.\  {\bf 48} (1988) 367.



\end{thebibliography}
\end{document}